\documentclass[twocolumn]{aastex631}

\shorttitle{EUV spectra and nanoflare flows}
\shortauthors{L\'opez Fuentes \& Klimchuk}

\graphicspath{{./}{figures/}}

\usepackage{mathrsfs}

\begin{document}

\title{The effect of nanoflare flows on EUV spectral lines}

\author[0000-0001-8830-4022]{Marcelo L\'opez Fuentes}
\affiliation{Instituto de Astronom\'{\i}a y F\'{\i}sica del Espacio (IAFE, CONICET-UBA), Buenos Aires, Argentina}

\author[0000-0003-2255-0305]{James A. Klimchuk}
\affiliation{NASA Goddard Space Flight Center, MD, USA}

\begin{abstract}

The nanoflare model of coronal heating is one of the most successful scenarios to explain, within a single framework, the diverse set of coronal observations available with the present instrument resolutions. The model is based on the idea that the coronal structure is formed by elementary magnetic strands which are tangled and twisted by the displacement of their photospheric footpoints by convective motions. These displacements inject magnetic stress between neighbor strands that promotes current sheet formation, reconnection, plasma heating, and possibly also particle acceleration. Among other features, the model predicts the ubiquitous presence of plasma flows at different temperatures. These flows should, in principle, produce measurable effects on observed spectral lines in the form of Doppler-shifts, line asymmetries and non-thermal broadenings. In this work we use a Two-Dimensional Cellular Automaton Model (2DCAM) developed in previous works, in combination with the Enthalpy Based Thermal Evolution of Loops (EBTEL) model, to analyze the effect of nanoflare heating on a set of known EUV spectral lines. We find that the complex combination of the emission from plasmas at different temperatures, densities and velocities, in simultaneously evolving unresolved strands, produces characteristic properties in the constructed synthetic lines, such as Doppler-shifts and non-thermal velocities up to tens of km s$^{-1}$ for the higher analyzed temperatures. Our results might prove useful to guide future modeling and observations, in particular, regarding the new generation of proposed instruments designed to diagnose plasmas in the 5 to 10 MK temperature range.

\end{abstract}



\section{Introduction}
\label{sect:intro}

One of the most popular theories to explain the high temperatures ($\gtrsim$ 1 MK) observed in the solar corona, particularly in active regions (ARs), is based on the idea that plasma heating is produced by flare-like mechanisms acting at sub-resolution level \citep{parker1988}. This kind of model explains many of the characteristics of coronal loops, the observable building blocks of ARs coronal structure \citep{reale2014}. AR loops are observed in a variety of wavelengths from EUV to X-rays. Although early studies of hot X-ray loops ($\sim$ 4 MK) suggested temperatures and densities that made them generally compatible with a regime of equilibrium between a quasi-steady heating source and losses produced by conduction and radiation \citep{rosner1978}, cooler EUV loops ($\sim$ 1 MK) were later observed to be too dense and dynamic to be consistent with that scenario \citep{aschwanden2001,winebarger2003}. Even hot loops may be incompatible, depending on the assumed filling factor \citep{klimchuk2015}.

The nanoflare model of coronal heating provides an evolutionary scheme that can explain the presence of both kinds of loops. In this scheme, loops are formed by sub-resolution elementary magnetic strands being heated by short duration impulsive events, or nanoflares, which are due to the sudden release of free magnetic energy associated with magnetic stress between neighbor strands \citep[for a recent review of Parker's problem, see ][]{pontin2020}. Ultimately, the magnetic stress comes from the constant shuffling and tangling of the strands' footpoints produced by photospheric convective motions. In this way, observed individual loops are the result of the simultaneous emission of bundles of strands each going through independent evolutions. In simple terms, the evolution of an individual strand being heated by an impulsive event such as a nanoflare can be separated in two main phases. In the first phase the plasma suffers a sudden increase of temperature during which the heat transport is dominated by thermal conduction. As temperature increases the energy imbalance with the transition region produces the evaporation of material into the coronal portion of the strand, increasing its density and therefore the capacity of the plasma to get rid of the excess thermal energy by radiation. In this second phase radiation cooling dominates. Later on, temperature and density slowly decrease towards previous equilibrium conditions (a model example of this evolution can be found in \citet[][see their Figure 3]{lopezfuentes2015b}. Depending on how often a nanoflare goes off in a given strand this evolution will be completed in full or in part. Therefore, the nanoflare frequency is a fundamental parameter in this scheme.

It has been argued that hot loops, such as those observed in AR cores, can be associated with high-frequency nanoflares \citep[see e.g., ][]{warren2011a}. In these loops individual strands never reach the cooling phase associated with high densities and with temperatures of the order of 1 MK. It is worth to note though, that recent works found observational evidence of nanoflares with short durations ($\approx$30 s) and low frequencies in hot loops of AR cores \citep[see, e.g., ][]{testa2014,ugarte-urra2019,testa2020}. If the nanoflare frequency is low enough, some or most of the strands may reach the low temperatures and high densities that explain the EUV observation of cooler loops at the periphery of AR cores (sometimes also called fan loops). In a series of works, \citet{viall2012,viall2013,viall2015} identified the presence of time lags in loop lightcurves obtained with different EUV channels of the Atmospheric Imager Assembly (AIA) on board SDO. They interpreted these lags as the evidence of a cooling process as the plasma went through the temperature sensitivity window of each wavelength channel. Further evidence of the presence of cooling processes have been found in other observations \citep[see e.g., ][]{terzo2011, lopezfuentes2016}. 

There are two key aspects of nanoflare heating that received special attention in recent years. One is the fact that if low-frequency nanoflares are a significant heating mechanism in loops, then there must be a very hot component ($\gtrsim$ 6 MK) of the plasma at all times. Direct and indirect observational evidence of the presence of this hot component has been found in recent works. \cite{brosius2014} using AR data from EUNIS-13, a sounding rocket mission, found pervasive emission in the Fe XIX line at 592.2~\AA, formed at T = 8.9 MK, supporting the predictions of the nanoflare model. Other evidence of high temperature in loops was found by, e.g., \citet{reale2009}, \citet{reale2011}, \citet{testa2012} \citep[see also references in][]{hinode-team2019}. 

Another important expected feature of nanoflare heating is the widespread presence of flows. The impulsive nature of the nanoflare scenario implies a dynamic exchange of plasma between the chromosphere/transition region and the corona. In the scheme described earlier in this Section heated strands will suffer both upflows and downflows at different stages of their evolution. If loops are made of collections of strands evolving independently, at any given time a mixture of both upflows and downflows are expected to be present below the resolution level. However, since upflows and downflows occur at different times during the evolution, and therefore at different characteristic temperatures, it could be ideally possible to identify separately the upflow and downflow contributions by analysing particular spectral lines of known formation temperatures. This can be a very complicated task as it was thoroughly described by \citet{young2012}. A series of studies based on spectroscopic observations \citep[][]{warren2011b,tripathi2012a,
tripathi2012b,ugarte-urra2012,winebarger2013,testa2014,testa2016,polito2018,testa2020}, provided flow velocity estimations for different temperature ranges. On the theoretical side, there has been a sustained effort to develop models based in different mechanisms and aproaches to forward model spectral diagnostic observations \cite[see e.g.,][]{peter2006,patsourakos2006,hansteen2010,pontin2020,depontieu2022}.

In two recent papers \citep[][ henceforth LFK15 and LFK16]{lopezfuentes2015,lopezfuentes2016}, we developed and studied a model based on the evolution of cellular automata that simulate the magnetic stressing and further reconnection between neighbor strands and the associated energy release in the form of nanoflares. To model the plasma response to the heating we used the Enthalpy Based Thermal Evolution of Loops model \citep[EBTEL, ][]{klimchuk2008, cargill2012}. In LFK16 we compared synthetic lightcurves obtained with the model with loop lightcurves from real EUV and X-ray observations. There, we also studied the emission measure (EM) distribution of the plasma obtained with the model and compared it with observational results from previous studies. Here, we revisit this Two-Dimensional Cellular Automaton Model (2DCAM) in combination with EBTEL (henceforth, 2DCAM-EBTEL) to analyze the predictions of the model regarding the effect of upflows and downflows on coronal spectral lines. To that end, we construct synthetic lines of known coronal ions by adding the emission contribution of the evolving plasma in sets of strands and individual nanoflares from the model. We then analyze the modeled lines following plasma diagnostic techniques usually applied to real spectral observations. The procedure is inspired by a similar approach previously followed by \citet{patsourakos2006}, with the main difference that here we use a model that provides a statistical sample of nanoflares in simultaneously evolving elemental strands. Also, here we analyze a larger number of spectral lines covering a wider temperature range (0.417 to 7.76 MK).

In Section~\ref{sect:model} we outline the implementation of the 2DCAM-EBTEL model for the present study, in Section~\ref{sect:lines} we describe the line construction and methods of analysis and in Section~\ref{sect:results} we present our results. We discuss our findings and conclude in Section~\ref{sect:conclusions}.


\section{Model description and implementation}
\label{sect:model}

\subsection{2DCAM-EBTEL}
\label{sect:2dcam-ebtel}

The layout of the two-dimensional cellular automaton model (2DCAM) developed in LFK15 consists of a set of moving points arranged in a two-dimensional square lattice. Initially, each lattice location is occupied by a single point. As the model evolves each point is randomly displaced to first neighbor locations one step at a time. This configuration simulates in a very crude way the displacements, produced by photospheric convective motions, of the footpoints of elementary magnetic strands that populate the solar corona. We relate the linear distance travelled by each point with an inclination of the associated magnetic strand and the consequent injection of magnetic stress. The mathematical details of these relations are thoroughly described in LFK15. Every time two points occupy the same location in the model lattice, we consider that the related strands are interacting. The model then considers the relative inclination of the strands (misalignment angle) and, if a given threshold is surpassed, the strands reconnect and free energy from the mutual magnetic stress is released in the form of a nanoflare. Then, the reconnected strands relax and the process continues. Despite its simplicity the model reproduces the main qualitative and numerical aspects of Parker's classical model. As shown in LFK15, the distributions of nanoflare energies produced by the model follow power-laws with a approximate mean index of -2.5, which is consistent with a scenario of coronal heating dominated by nanoflares \citep{hudson1991}.

The input parameters of 2DCAM are: the number of strands, $N_s$; the strand length, $L$; the model time-step duration, $\delta t$, which we associate with the photospheric convective turnover time; the mesh parameter, $D$, i.e., the distance between adjacent points in the lattice that we associate with a typical convective granule size; and the threshold inclination between interacting strands, $\theta_c$. The magnetic strength assigned to the strands is obtained from a relation between magnetic field strength and loop length found by \citet{mandrini2000}. The effects of the variations of these parameters on the model's output have been thoroughly studied in LFK15 and LFK16. Here, we use mean characteristic values which are typical of solar active regions: $L~=$~100~Mm, $\delta t=$~1000~s, and $D=$~1000~km. For the number of strands and the threshold inclination angle we use intermediate reasonable values of $N_s=$~49 and $\theta_c =$~14~deg.

The 2DCAM model provides the energies of the nanoflares going off on each strand as a function of time. To model the response of the plasma in the individual strands to the heating, we use the 0D Enthalpy Based Thermal Evolution of Loops model \citep[EBTEL, ][]{klimchuk2008, cargill2012}). EBTEL provides, as a function of time, the spatially averaged temperature, density and differential emission measure (DEM) along the coronal portion of the strand, the velocity at the coronal base, and the transition region (TR) DEM. As input for EBTEL we model each nanoflare in the individual strands as triangular heating functions of 200 s durations and 1 s temporal resolution. This nanoflare duration provided a good agreement with observed loop intensities and their fluctuations in our previous works (LFK15, LFK16). Recently, Klimchuk, Knizhnik \& Uritsky (2022, \textit{ApJ}, submitted) also found that nanoflare durations of 500 s or less are consistent with their MHD simulations. We postpone the analysis of the variation of the nanoflare duration parameter for a future study. At this level of modeling we decided to avoid adding another variable parameter to the analysis. The final output is a series of arrays containing the temporal evolution of the described mean coronal and TR plasma parameters for each of the modeled magnetic strands. The model run used  in this study has a total duration of $10^5$ s. In Section~\ref{sect:lines} we describe how we construct synthetic spectral lines from this output.

\subsection{Example of single-strand evolution}
\label{sect:single_strand}

\begin{figure*}[ht!]
\centering
\hspace{0.cm}
\includegraphics[width=0.6\textwidth]{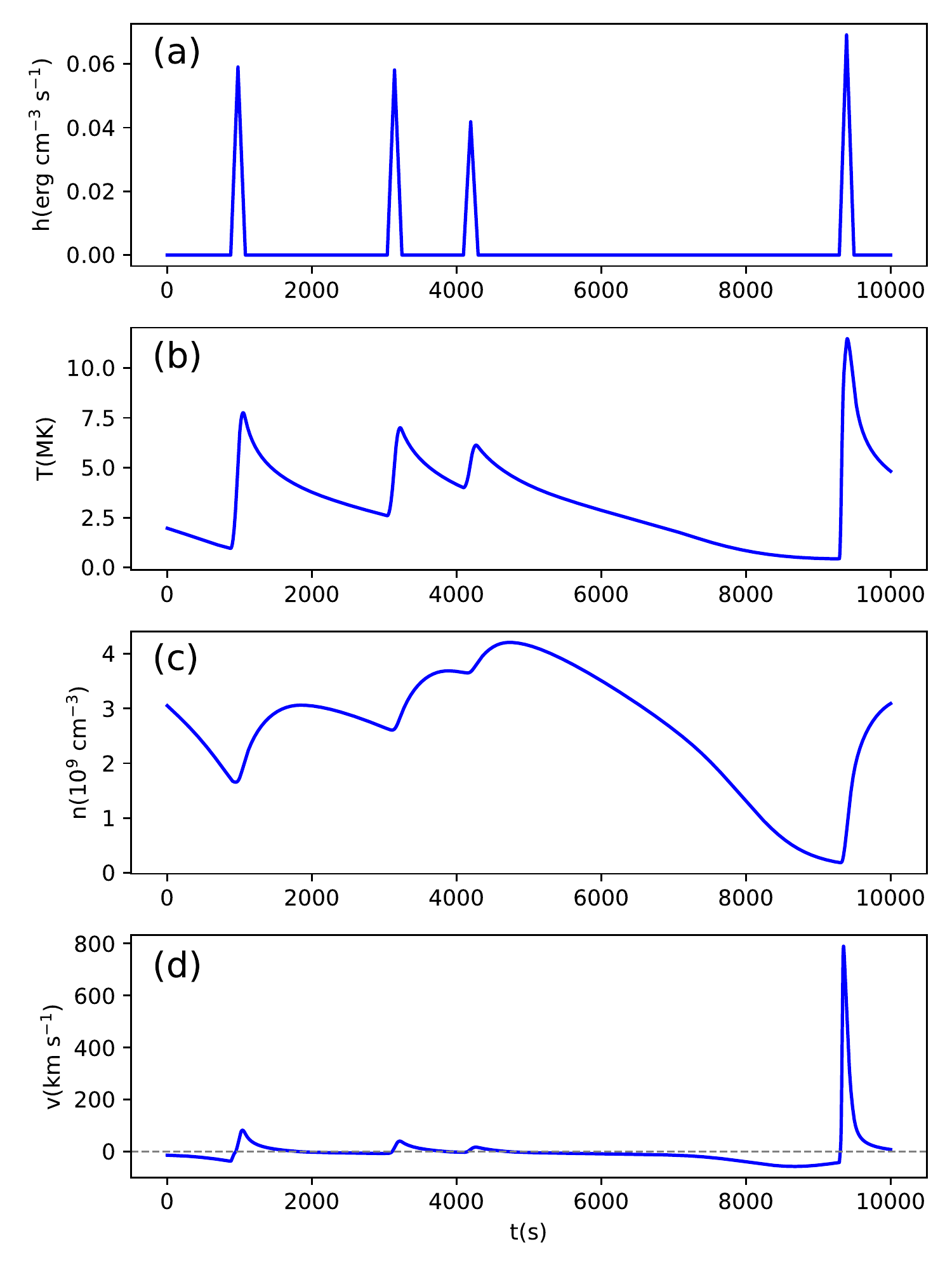}
       \caption{Example of evolution along 10$^{4}$ s of a single strand extracted from the 2DCAM-EBTEL model. Panel a: Sequence of 200 s triangular nanoflares that heat the plasma. Panel b: Evolution of the mean coronal temperature of the plasma computed with EBTEL. c) Idem mean coronal density. Panel d: Idem plasma velocity.}
\label{fig:evolution}
\end{figure*}

To further illustrate how the model works, in Figure~\ref{fig:evolution} we show the evolution of the mean coronal plasma parameters of a single strand for a selected time interval of 10$^4$ s. Figure~\ref{fig:evolution}a shows the evolution of the heating rate where four nanoflares of different energies can be appreciated. The model has also a constant background low intensity heating component of 10$^{-5}$ erg cm$^{-3}$ s$^{-1}$. Panel b shows the temperature evolution where the sudden increases of temperature at the times when the nanoflares turn on are seen. As soon as the heating goes down the temperature begins to fall until the next nanoflare occurs. Panel c corresponds to the mean coronal density of the strand. It can be seen that whenever the nanoflares occur and the temperature increases the density begins to rise as the strand gets filled with plasma from the chromosphere and transition region. The subsequent cooling from radiation and thermal conduction decreases the temperature and the pressure support, and the strand drains plasma to the lower layers. Finally, panel d shows the mean plasma velocity in the coronal portion of the strand, where positive velocities (upflows) correspond to the times when the strand is being filled (increasing density) and negative velocities (downflows) when plasma drains down (decreasing density).  

The example shown in Figure~\ref{fig:evolution} was selected because it shows typical nanoflare energies for the combination of model parameters used here (between 4 and 7 erg cm$^{-3}$, see also LFK15) and three different cases of waiting times between consecutive nanoflares (or nanoflare frequencies). As we thoroughly explain in Section~\ref{sect:nanoflares}, the time difference between the first and the second nanoflares of Figure~\ref{fig:evolution}a, $\sim$2000 s, corresponds to a case of intermediate frequency, the difference between the second and the third, $\sim$1000 s, to high frequency, and the time between the third and the fourth nanoflare, $\sim$5000 s, to low frequency. Note that in the last case the temperature and density reach very low values, so when the fourth nanoflare suddenly goes off, fast chromospheric evaporation produces extremely high upflow velocities which make the previous upflow events look comparatively small. As we show in Section~\ref{sect:results}, these high velocities are smeared out in the synthetic spectra due to the emission weighting produced by the integrated contributions of strands at different evolutionary stages. In Section~\ref{sect:results} we use the example of Figure~\ref{fig:evolution} as one of the test cases for how the plasma temperature, density and velocity evolution affect modeled spectral lines. 

\subsection{Individual nanoflares}
\label{sect:nanoflares}
	
The 2DCAM-EBTEL model decribed in Section~\ref{sect:2dcam-ebtel} produces nanoflares with a variety of characteristics that define the evolution of the plasma in the magnetic strands. In Section 4 we analyze spectral lines constructed from the combined emission of these strands. To help us understand how the strands' evolution produces particular features of the modeled lines, we analyze the effect that individual nanoflares with different characteristics have on the lines. In this Section we describe how we model those nanoflares and the parameters we use. The three main characteristics of nanoflares that we consider in our analysis are: the total nanoflare energy, the nanoflare frequency, and the initial conditions at the time when the nanoflare starts. 

Although the nanoflare energy distributions produced by the 2DCAM model follow power laws with an approximate index of -2.5, the actual energy ranges in the distribution depend primarily on the loop length. For the loop length used in this work ($L=$100 Mm) nanoflare energies range between 1 and 25 erg cm$^{-3}$, with a typical mean value of 5 erg cm$^{-3}$. Thus, we use this energy as the reference value on which we perform the variation of the different parameters. 

\begin{figure*}[ht!]
\centering
\hspace{0.cm}
\includegraphics[width=0.6\textwidth]{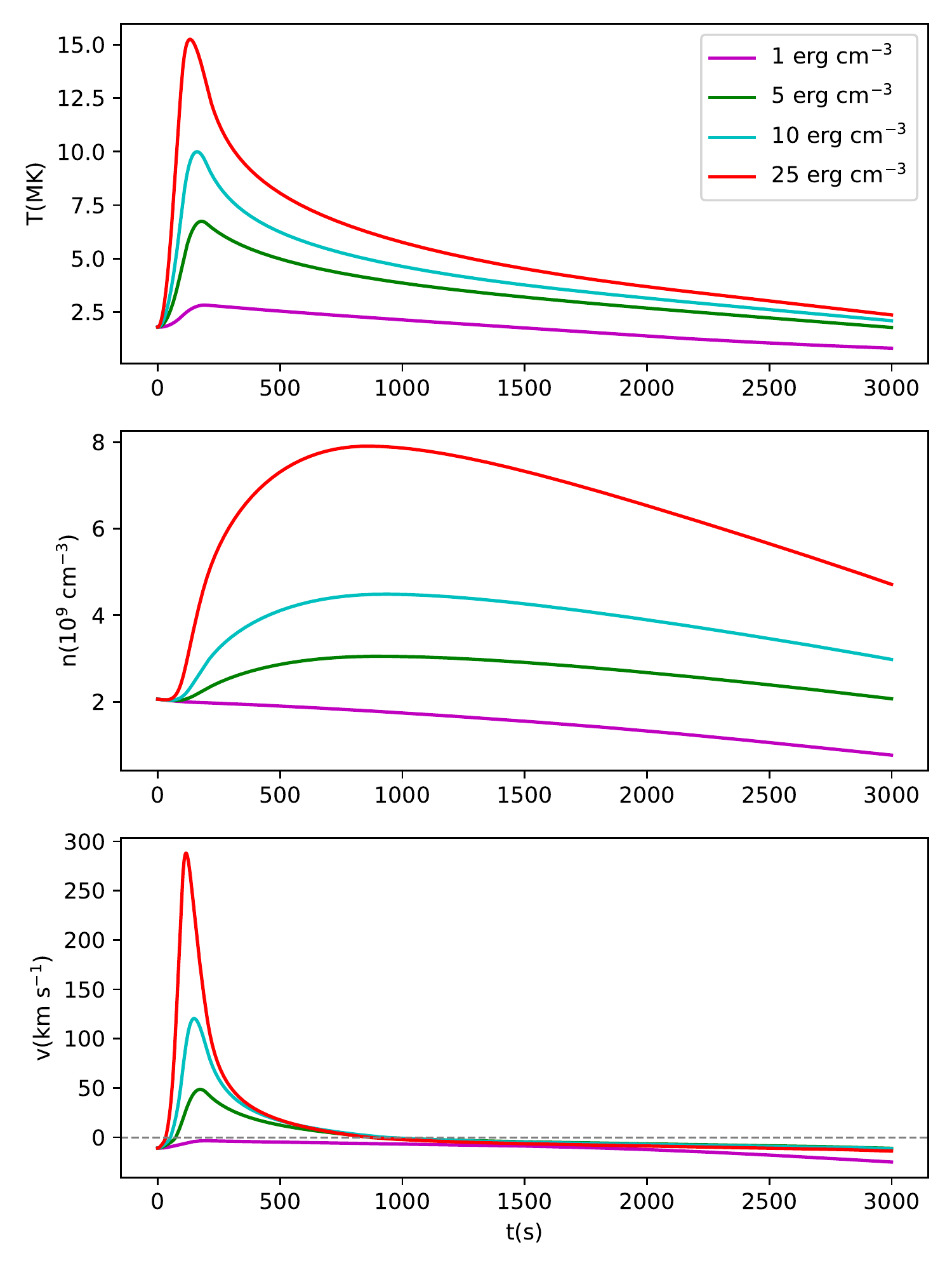}
       \caption{Evolution of single 200 s nanoflares with different volumetric energies computed with EBTEL for a strand of 100 Mm. Top panel: mean coronal temperature. Middle panel: mean coronal density. Bottom panel: mean coronal velocity.}
\label{fig:energy}
\end{figure*}

In order to analyze how the nanoflare energy affects the characteristics of the modeled spectral lines, we use EBTEL to obtain the evolution of four individual nanoflares with total volumetric energies of 1, 5, 10 and 25 erg cm$^{-3}$. In Figure~\ref{fig:energy} we plot the evolution of the mean coronal temperature and density and the velocity of these nanoflares. The heating is applied as triangular functions of 200 s durations with maximum rates of 0.01, 0.05, 0.1 and 0.25 erg cm$^{-3}$ s$^{-1}$, respectively. As identical initial conditions for all these nanoflares we used $T=$ 1.8 MK, $n=2\times 10^9$ cm$^{-3}$ and $v=-10^6$ cm s$^{-1}$ (downflow), which can be considered typical as we explain later in this Section. Notice in the corresponding panels of Figure~\ref{fig:energy} how the increasing nanoflare energy increases the maximum temperature, density and velocity of the plasma. As we analyze in Section~\ref{sect:results}, the variation of the thermal parameters with the nanoflare energy produces characteristic signatures on the modeled spectral lines.

The nanoflare frequency - the rate at which nanoflares recur in a given strand - affects the plasma evolution in two ways. First, it determines how hot the plasma gets and therefore how strong the ensuing evaporation is. A relatively dense plasma is heated to a lower peak temperature by a given nanoflare energy release. Second, it determines how much the plasma cools before being reheated by the next event. We consider each of these effects in turn, starting with the second. We use the same classification of nanoflare frequency as adopted in LFK15, which is based on the relation between the repetition time between consecutive nanoflares and the characteristic cooling times after the plasma reached the maximum temperature. According to these definitions, the nanoflare frequency is considered high if the repetition time is shorter than the time required for the temperature to decrease to 61\% of its maximum value, intermediate frequencies correspond to repetition times for which the temperature reaches between 14\% and 61\% of its maximum, and low frequencies correspond to the case of the temperature reaching less than 14\% of its maximum value. Considering for the present model a typical nanoflare of 5 erg cm$^{-3}$, the repetition times are approximately 1000 s for high nanoflare frequencies, between 1000 s and 4000 s for intermediate frequencies and more than 4000 s for low frequencies. The strand evolution of the example shown in Figure~\ref{fig:evolution} illustrates approximately the three described situations: high frequency between the second and the third nanoflare, intermediate frequency between the first and the second nanoflare and low frequency between the third and the fourth nanoflare. To probe the effect of the repetition times (equivalently, nanoflare frequencies) on the amount of cooling that takes place, and the consequences it has for the modeled spectral lines, we integrate the emission from the 5 erg cm$^{-3}$ nanoflare simulation for the first 1000 s (high frequency), the first 3000 s (intermediate frequency) and the first 5000 s (low frequency). It is worth mentioning here that, as we found in LFK15 (see our Table 3 there), for the 2DCAM-EBTEL model with 100 Mm strands, 43\% of the nanoflares correspond to high frequency, 50\% to middle frequency and only 7\% to low frequency. As we see in the following sections, this has consecuences on the measured properties of the spectral lines modeled with the 2DCAM-EBTEL model. 

\begin{figure*}[ht!]
\centering
\hspace{0.cm}
\includegraphics[width=0.6\textwidth]{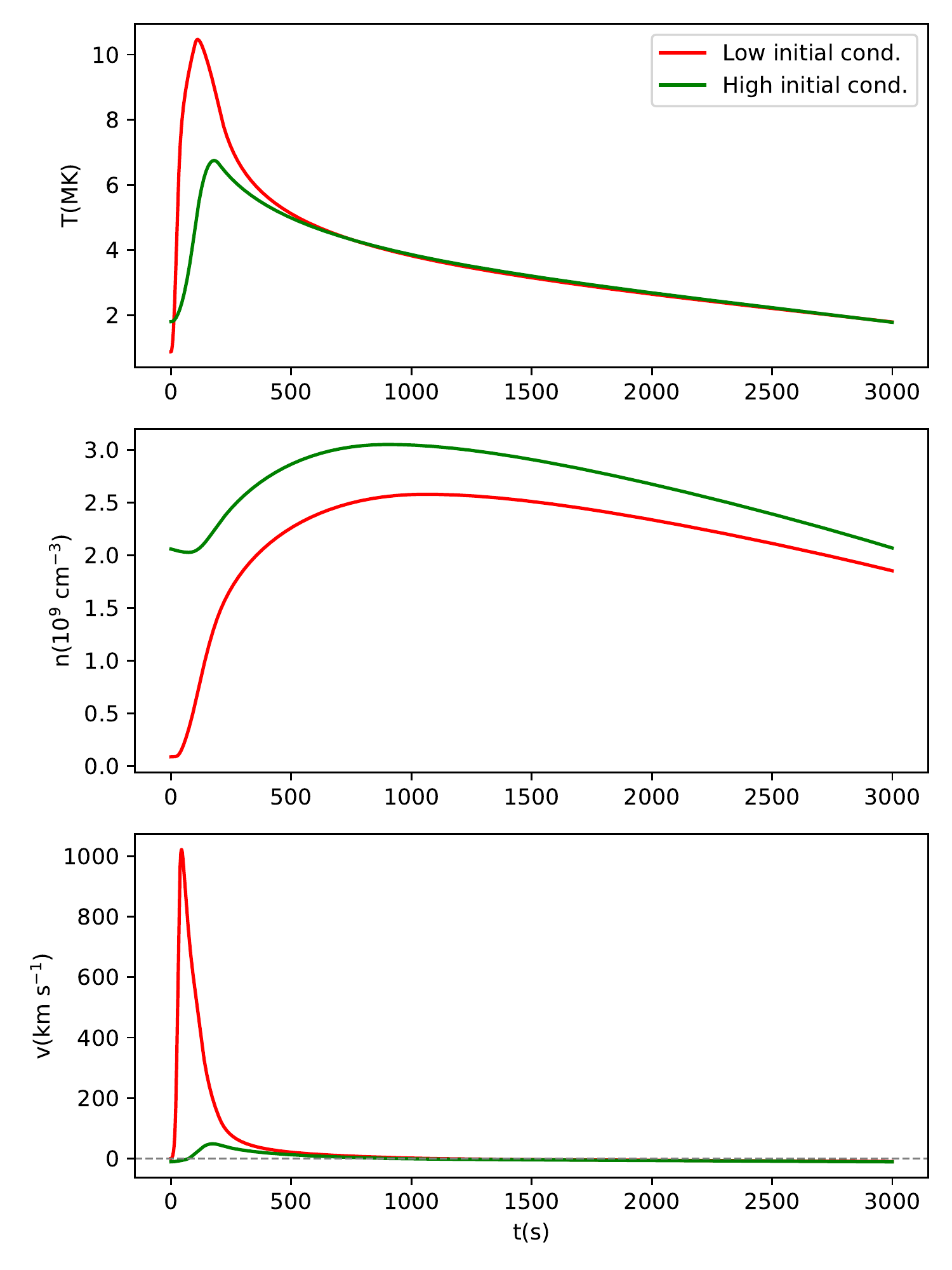}
       \caption{Idem Figure~\ref{fig:energy} for 5 erg cm$^{-3}$ nanoflares having different initial conditions. Low initial conditions: $T=$ 0.88 MK, $n=9\times10^{7}$ cm$^{-3}$, $v=0$. High initial conditions: $T=$ 1.8 MK, $n=2\times10^{9}$ cm$^{-3}$, $v=-10^{6}$ cm s$^{-1}$.}
\label{fig:init_cond}
\end{figure*}

As we have indicated, and as can be seen in Figure~\ref{fig:evolution}, the nanoflare frequency also determines the physical conditions at the start of each event, which impacts the subsequent evolution (compare for example the initial conditions of the third and fourth nanoflares). To study the effect of the initial conditions on the evolution of a single nanoflare, and on the modeled spectral lines, we model two nanoflares of the same energy (5 erg cm$^{-3}$) with initial conditions corresponding to a case of intermediate frequency and a case of low frequency. Figure~\ref{fig:init_cond} shows the evolution of these two cases. For the low frequency initial conditions case we use initial equilibrium conditions provided by EBTEL, $T=$ 0.88 MK, $n=9\times 10^7$ cm$^{-3}$ and $v=$ 0. The corresponding evolution is represented with red lines in the panels of Figure~\ref{fig:init_cond}. Since most nanoflares in the model are preceded by other nanoflares, these are not the initial conditions in most cases. To simulate more regular initial conditions (identified as ``high initial conditions'' in the legend of Figure~\ref{fig:init_cond}) we run three consecutive nanoflares of 5 erg cm$^{-3}$, separated by 3000 s, corresponding to an intermediate frequency, and we use the third one as the nanoflare of interest. We found that starting from the third nanoflare, if we continue adding events, the evolutions of the following cases are nearly exactly the same. We therefore consider the third nanoflare the generic case for a typical nanoflare energy and a standard intermediate frequency in our model. The corresponding generic initial conditions in this case are $T=$ 1.8 MK, $n=2\times 10^9$ cm$^{-3}$ and $v= -10^6$ cm s$^{-1}$. These are the initial conditions that we chose to use in the different nanoflare energy runs of Figure~\ref{fig:energy}. 

\section{Synthetic spectral lines construction and analysis}
\label{sect:lines}

To explore the effect that flows produced by nanoflares have on EUV spectral lines we use the thermal and velocity properties of the plasma in the evolving modeled strands in the following way. We compute the instantaneous contribution of strand $j$ to the emission of ion line $i$ as

\begin{equation}
I_{ij}(t) = n_j^2(t) G_i(T_j(t)),
\label{eq:int}
\end{equation}

\noindent where $n_j(t)$ is the strand density, $G_i(T)$ is the contribution function of the ion $i$ obtained from the CHIANTI database \citep{delzanna2015} and $T_j(t)$ is the mean coronal temperature of the strand. We model the line using a gaussian profile:

\begin{equation}
I(\lambda) = A \exp{\left(-\frac{(\lambda-\lambda_j)^2}{w_j^2}\right)},
\label{eq:profile}
\end{equation}

\noindent centered at the Doppler shifted wavelength:

\begin{equation}
\lambda_j(t) = \lambda_0 \left(1 - \frac{V_j(t)}{c}\right),
\label{eq:doppler}
\end{equation}

\noindent where $V_j(t)$ is the plasma velocity (positive for upflow), $c$ is the speed of light and $\lambda_0$ is the line central wavelength for the ion at rest ($V=0$). The amplitude $A$ is defined so that $I_{ij} = \int I(\lambda) d\lambda$ and $w_j$ corresponds to the $1/e$ half-width:   

\begin{equation}
w_j(t) = \frac{\lambda_0}{c}\left(\frac{2kT_j(t)}{m_i}\right)^{\frac{1}{2}}.
\label{eq:width}
\end{equation}

\noindent In the previous equation, $k$ is the Boltzmann constant, $m_i$ is the ion mass, and $T_j$ corresponds to the ion temperature of the strand, which we take equal to the electron temperature in all our analysis. For the plasma temperature, density and velocity in equations~\ref{eq:int} and~\ref{eq:doppler} we use the mean coronal values provided by EBTEL. 

The procedure just described provides the coronal contribution of the modeled spectral line and it does not include the transition region (TR) emission. Since the ranges of temperature and density are so great in the TR of each strand, it is not appropriate to use mean values. Instead, we compute spectral line intensities using the differential emission measure distribution (DEM), which is an EBTEL output. The intensity contribution is expressed in differential form as

\begin{equation}
\mathrm{d}I_{TR} = DEM_{TR}(T) G(T) \mathrm{d}T,
\label{eq:dinttr}
\end{equation}

\noindent where $DEM_{TR}(T)$ is the DEM of the TR plasma provided by EBTEL and $G(T)$ is the ion contribution function, as before. In Equation~\ref{eq:dinttr} we removed the $i$ and $j$ indeces for simplicity. The instantaneous line profile is obtained using Equations~\ref{eq:profile} and~\ref{eq:doppler}, similarly to the coronal case, but integrating it with the differential intensity from Equation~\ref{eq:dinttr} between $T_b = 0.3$ MK and $T_0$, which are, respectively, the minimum relevant temperature for the present analysis and the temperature at the top of the TR (i.e., the coronal base of the strand) obtained with EBTEL. The TR plasma velocity used to compute the Doppler shift (Equation~\ref{eq:doppler}) included in the line integration is:

\begin{equation}
V_{TR}(t) = 1.5  \frac{T}{T_j(t)} V_j(t),
\label{eq:vtr}
\end{equation}

\noindent where $T_j$ and $V_j$ are the mean coronal temperature and the velocity at the coronal base of the strand (as in Equations~\ref{eq:profile} and~\ref{eq:doppler}), and $T$ is the TR temperature, i.e., the integration variable of Equation~\ref{eq:dinttr}. The expression of Equation~\ref{eq:vtr} derives from \citet[][see their Equation 24]{klimchuk2008}. The factor of 1.5 is the ratio of the average coronal temperature to the temperature at the coronal base ($T_{0}$). It is worth to add here that EBTEL defines the boundary between the corona and TR to be the place where thermal conduction switches from a cooling term above to a heating term below. This occurs at a temperature that is roughly 60\% of the apex temperature of the strand. The coronal velocity is taken to be the velocity at this boundary. The velocity at any temperature $T$ in the transition region is obtained with the reasonably assumptions of constant mass flux at constant pressure.

If we consider observations on the solar disk, to put the coronal and TR contributions on equal foot it is necessary to include in the computation the depth of the plasma column contributing to the emission. The computation of $DEM_{TR}$ made by EBTEL includes the TR column depth, but the coronal intensity obtained from Equation~\ref{eq:int} corresponds to the emission per unit volume. To include the effect of the coronal column depth we follow the same approach as \citet{klimchuk2014b}, which considers that the emission along a single magnetic strand is representative of the emission through an arcade of strands that lie along the line of sight (see their Figure 1). Following this approximation, in order to account for the emission of the coronal plasma column we multiply the line intensity obtained from Equation~\ref{eq:int} by the strand half-length.

The instantaneous line profile produced by the strand, including the coronal and TR contributions, is then integrated over a given interval of time. In the case of individual nanoflares (see Section \ref{sect:nanoflares}) we integrate the line over the corresponding evolution and in the case of the single strand example of Figure \ref{fig:evolution} the integration is done over the full 10$^{4}$ s evolution shown. In the case of synthetic spectral lines constucted by adding the contribution of all the strands in the 2DCAM-EBTEL model, we use intervals of 30 s, corresponding to the typical exposure time of observations performed, for instance, with the EUV Imaging Spectrometer (EIS) on board Hinode. In this work we study five such 30 s intervals, or samplings, selected from the full evolution of the model ($10^5$ s), starting at $t = 3\times 10^4$ s, to be sure that the system is fully developed, and separated from each other by $10^4$ s. The $10^4$ s separation between samplings guarantees that they are completely independent.

Although the main analysis of this paper concerns the output of the 2DCAM-EBTEL model integrated in the five 30 s samplings described, the line integrations for individual nanoflares described in Section 2.3 and the individual strand example described in Section 2.2, would be valuable to help us understand how different nanoflare characteristics, such as the heating rate or the nanoflare frequency, affect the measurable properties of the modeled spectral lines (see Section 4).

In Table~\ref{table1} we list the 11 spectral lines used in the present analysis. The first column indicates the ion, the second column provides the line wavelength and the third column shows the line formation temperature. The content of the rest of the columns is described in the following sections. 

\begin{table*}
\centering
\caption{Properties of modeled spectral lines. The corresponding columns contain from left to right: Ion, line formation wavelength ($\lambda_0$), line formation temperature ($T$), mean value of Doppler-shift velocities computed from lines integrated using 2DCAM-EBTEL model samplings ($V_D$), corresponding standard deviations of the Doppler-shift velocity means ($\sigma_{V_D}$), mean of the TR to coronal intensity ratio of the analyzed lines, mean asymmetry parameter ($\mathscr{A}$, see definition in Section~\ref{sect:asym}), mean thermal velocity ($V_0$), mean non-thermal velocity ($\xi$) and its standard deviation ($\sigma_{\xi}$) (see text for detailed explanations).}
\begin{tabular}{lcccccccccc}
Ion	& $\lambda_0$[\AA] & $T$[MK] &  $V_D^{(*)}$ & $\sigma_{V_D}^{(*)}$ & $I_{TR}/I_{cor}$ & $\mathscr{A}$ & $V_0^{(*)}$ & $\xi^{(*)}$ & $\sigma_{\xi}^{*}$\\
\hline
FeVIII	&	194.66	&	0.417	&	0.87		&	0.35		&	452.95	&	0.022	    &	11.12       &	9.44   	& 	0.82 \\
SiVII	&	275.35	&	0.589	&	0.99		&	0.52		&	256.24	&	0.014	    &	18.60       &	9.77		&	0.77 \\
FeX	    &	184.54	&	0.977	&	1.28		&	0.41		&	87.05	&	0.012	    &	17.03	    &	8.74 	&  	1.31 \\
FeXI		&	188.23	&	1.17		&	1.31		&	0.58		&	61.93	&	0.008	    &	18.63	    &	9.68		& 	1.26 \\
FeXII	&	195.12	&	1.38		&	1.54		&	0.52		&	26.00	&	0.006	    &	20.24	    &	10.47	&	1.32	 \\
FeXIII	&	202.04	&	1.58		&	1.65		&	0.52		&	15.13	&	$3\times 10^{-4}$ &	21.65	&	11.39	&	1.42 \\
FeXIV	&	274.2	&	1.82		&	1.72		&	0.29		&	7.22		&	-0.008		&	23.24		&	12.63	&	1.36 \\
FeXV		&	284.16	&	2.09		&	1.57		&	0.77		&	2.76		&	-0.010		&	24.90		&	15.50	&	1.17	 \\
CaXIV	&	193.87	&	2.95		&	1.57		&	1.09		&	0.89		&	-0.011		&	34.86		&	16.03	&	2.20 \\
FeXVII	&	254.35	&	5.37		&	-1.83	&	1.73		&	0.17		&	-0.036		&	39.92		&	12.16	&	9.66 \\
FeXIX	&	592.24	&	7.76		&	-14.04	&	6.82		&	0.03		&	-0.104		&	47.99		&	30.17	&	21.30 \\
\hline
\footnotesize{$^{(*)}[$km s$^{-1}]$}
\end{tabular}\\
\label{table1}
\end{table*}

\begin{figure*}[ht!]
\centering
\hspace{0.cm}
\includegraphics[width=0.6\textwidth]{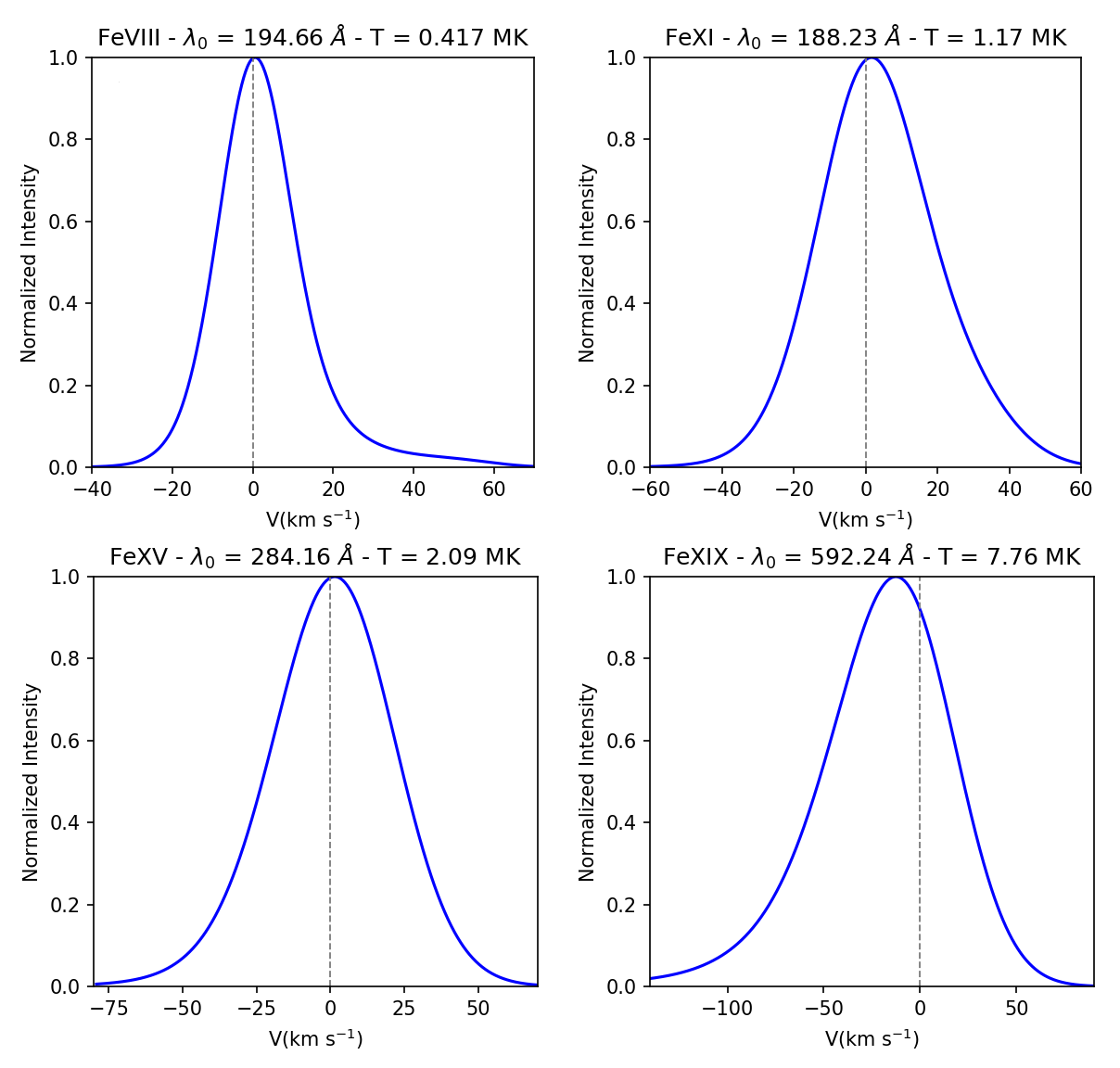}
       \caption{Examples of spectral line profiles integrated from the evolution of the 2DCAM-EBTEL strand shown in Figure~\ref{fig:evolution}. The corresponding ions, wavelengths and formation temperatures are shown on top of the panels. The line intensities are normalized and the original wavelengths at the abscisas were transformed into velocities by inverting Equation~\ref{eq:doppler}.}
\label{fig:examples}
\end{figure*}

To illustrate the effect that diverse evolutions have on the modeled spectral lines, in Figure~\ref{fig:examples} we show four examples of lines integrated along the single strand 10$^4$ s evolution described in Section~\ref{sect:single_strand} and shown in Figure~\ref{fig:evolution}. The line profiles shown correspond to the ions: FeVIII (panel a), FeXI (panel b), FeXV (panel c) and FeXIX (panel d). The profiles are normalized and plotted as a function of velocity, according to the relation between wavelength and velocity obtained by inverting the expression in Equation~\ref{eq:doppler}, as it is usually done in this kind of analysis. When velocity is used to represent Doppler shift, a negative value corresponds to blue shift, or upflow. When discussing the physical velocity of the plasma in the simulation, a negative velocity is a downflow. The vertical dashed lines indicate the expected positions of the spectral line centers for the ion at rest. A visual inspection of the profiles shows that they  are, in different degrees, shifted from their expected central positions and asymmetrical regarding their shape. These shifts and deformations are a consequence of the varying temperature, density and velocity of the modeled strand during the different stages of its evolution. The line profiles shown in Figure~\ref{fig:examples} resemble those obtained by \citet{patsourakos2006}, that applied a different model for the plasma evolution in their study based on individual nanoflares.

It is worth noting at this point that, since the characteristics of the constructed spectral lines are produced by the mixing and smearing of the summed effects of plasma at different evolutionary stages, these characteristics are comparable with actual observations only in a statistical fashion, assuming that the parameters chosen for the 2DCAM-EBTEL model correspond to a reasonable representation of the solar active regions case. As the model actually reproduces other observed features of the coronal plasma (as shown in LFK16), we assume that some of the approximations made would not affect the qualitative conclusions of the present analysis. It is also reasonable to expect that the quantitative aspects of the comparison with observations will be useful to guide future investigations.

In order to study the characteristics of spectral lines predicted by the nanoflare model we analyze the lines using diagnostic techniques commonly applied to real observations. For instance, the first moment (i.e., the centroid or mean value) of the intensity profile of the modeled line,

\begin{equation}
M_1 = \frac{\int{\lambda I(\lambda) d\lambda}}{\int{I(\lambda) d\lambda}},
\label{eq:1stmom}
\end{equation}

\noindent can be associated with a mean Doppler shift velocity, $V_D$, defined as

\begin{equation}
V_D = \left(\frac{M_1}{\lambda_0} - 1 \right) c.
\label{eq:dopvel}
\end{equation}

Similarly, the second moment (i.e., the variance) of the line profile, 

\begin{equation}
M_2 = \frac{\int{(\lambda - M_1)^2 I(\lambda) d\lambda}}{\int{I(\lambda) d\lambda}}
\label{eq:2ndmom}
\end{equation}

\noindent can be associated with the line broadening produced by non-thermal processes that translate into a velocity component, $\xi$, according to equation

\begin{equation}
M_2 = \frac{\lambda^2}{2c^2}\left(\frac{2kT}{m_i}+\xi^2\right),
\label{eq:nontherm}
\end{equation}

\noindent where

\begin{equation}
V_0 = \left(\frac{2kT}{m_i}\right)^{\frac{1}{2}}
\label{eq:therm}
\end{equation}

\noindent is the thermal velocity component, defined as the 1/$e$ half-width of a Maxwellian line profile \citep{dere1993}. The non-thermal velocity component is commonly associated with processes such as oscillations, turbulence or, as is the case here, spatially unresolved plasma flows \citep[see e.g., ][]{patsourakos2006,hahn2014}.


\section{Results}
\label{sect:results}

\subsection{EM weighted mean velocity}
\label{sect:demv}

\begin{figure*}[ht!]
\centering
\hspace{0.cm}
\includegraphics[width=0.6\textwidth]{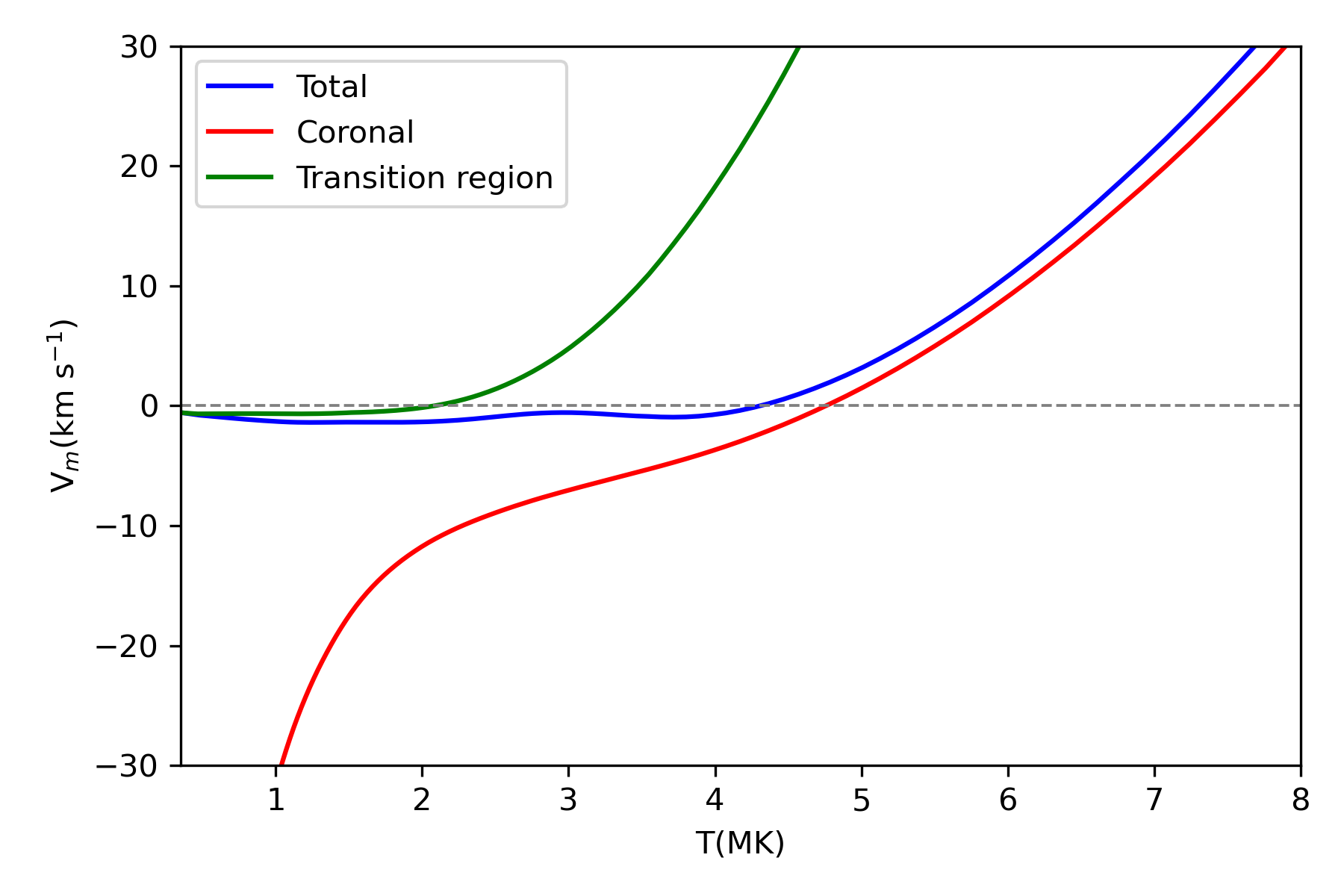}
       \caption{DEM weighted mean velocity of the plasma for the full 2DCAM-EBTEL model used here. The curve labelled "Total" corresponds to the expression of Equation~\ref{eq:demv} computed including both, the TR and coronal contributions, and the curves labelled "Coronal" and "Transition region" include only the corresponding contributions.}
\label{fig:demv}
\end{figure*}

Before analyzing the characteristics of the modeled spectral lines, let us study the relative statistical weight of the different velocity contributions to the plasma emission expected from the 2DCAM-EBTEL model. This will be useful as an initial guide for the analysis of the modeled lines. Let us define the EM weighted mean velocity as a function of temperature:

\begin{equation}
V_{m}(T) = \frac{\int{DEM(T) V(T) dt}}{\int{DEM(T) dt}},
\label{eq:demv}
\end{equation}

\noindent where $DEM(T)$ and $V(T)$ are the instantaneous DEM and velocity of the plasma on each strand. The integrations are done over all strands and the full evolution of the model (10$^5$ s). Since the EBTEL model provides the coronal and TR outputs separately, the integrals in the above expression are obtained summing both contributions. In Figure~\ref{fig:demv} we plot $V_{m}(T)$ computed considering the total sum of the coronal and TR contributions to the integrals in the numerator and denominator of the right hand of Equation~\ref{eq:demv} (blue curve), and the same computations considering, alternatively, only the coronal (red curve) and only the TR (green curve) contributions. We focus on the temperature range relevant for the present analysis. Since the coronal and TR DEM distributions diminish very rapidly at the boundaries of that range, according to its definition from Equation~\ref{eq:demv}, $V_m(T)$ diverges or converges to zero near those boundaries, as can be readily observed from the curves in Figure~\ref{fig:demv}. There, we use the sign convention of the EBTEL model, with positive values corresponding to velocities outward from the photosphere, and therefore towards the observer in on-disk observations. That means that in Figure~\ref{fig:demv} positive velocities correspond to upflows and negative velocities to downflows. The blue curve in Figure~\ref{fig:demv} shows that the total $V_m(T)$ has a dominancy of relatively small velocity downflows for temperatures up to approximately 4.3 MK and rapidly increasing velocity upflows for larger temperatures. A comparison with the TR (green) and the coronal (red) curves indicates that the lower temperature downflows are dominated by the TR contribution and the higher temperature upflows are dominated by the coronal contribution. We confirm this with the analysis of the modeled spectral lines of the next sections.

\subsection{Doppler-shift velocity}
\label{sect:doppler}

\begin{figure*}[ht!]
\centering
\hspace{0.cm}
\includegraphics[width=0.6\textwidth]{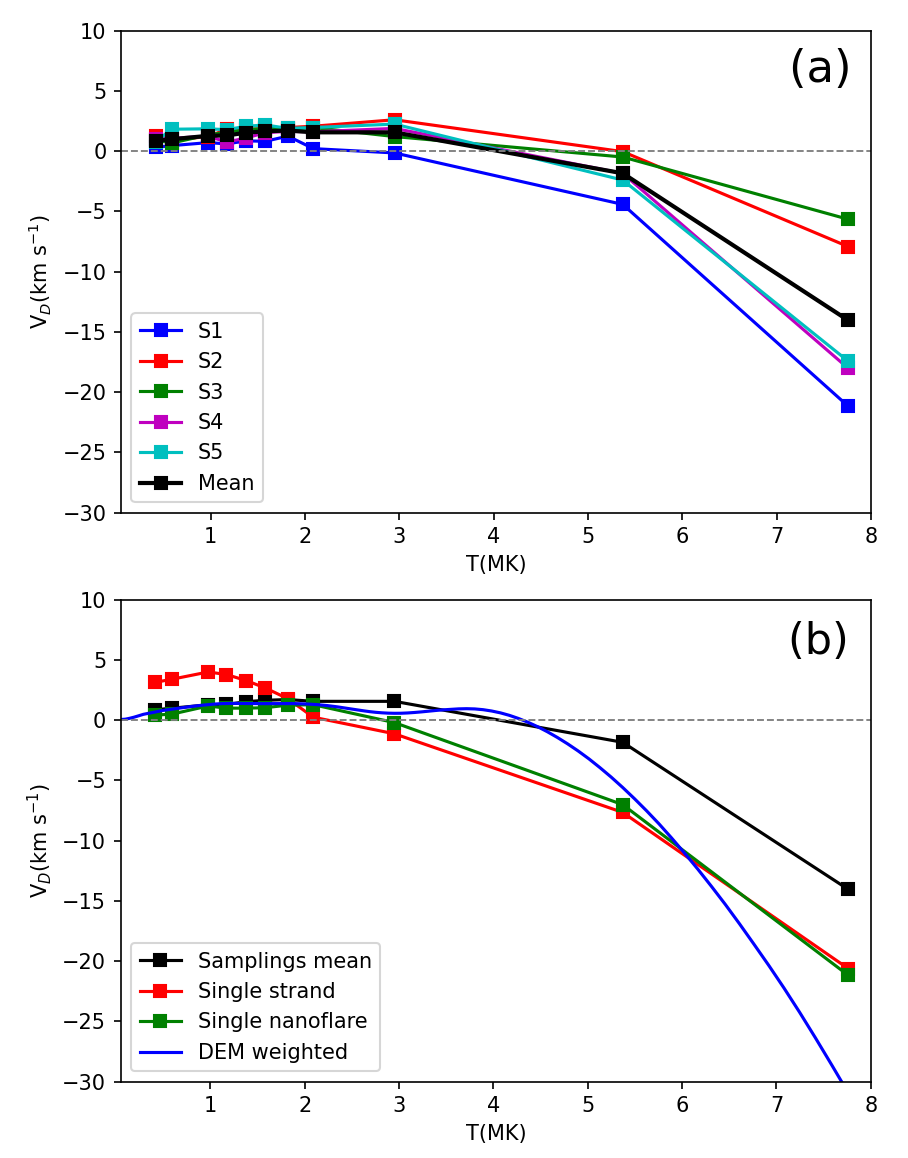}
       \caption{Doppler-shift velocities, as a function of temperature, of spectral line profiles integrated from the evolution of single and multiple modeled strands. Panel a: Five 30 s samplings from the 2DCAM-EBTEL model. The black squares and lines correspond to the mean values. Panel b: Comparison of the mean of panel a (the black curve) with the Doppler-shifts of lines integrated from the 10$^{4}$ s single strand evolution shown in Figure~\ref{fig:evolution} (red curve) and the single 5 erg cm$^{-3}$ standard nanoflare evolution of Figure~\ref{fig:energy} (green curve). Also included is the DEM weighted mean velocity of the full 2DCAM-EBTEL model shown in Figure~\ref{fig:demv} (blue curve). The latter is inverted with respect to Figure~\ref{fig:demv} because of the change of sign to the observer's point of view.}
\label{fig:dop_vel_loop}
\end{figure*}

\begin{figure*}[ht!]
\centering
\hspace{0.cm}
\includegraphics[width=0.6\textwidth]{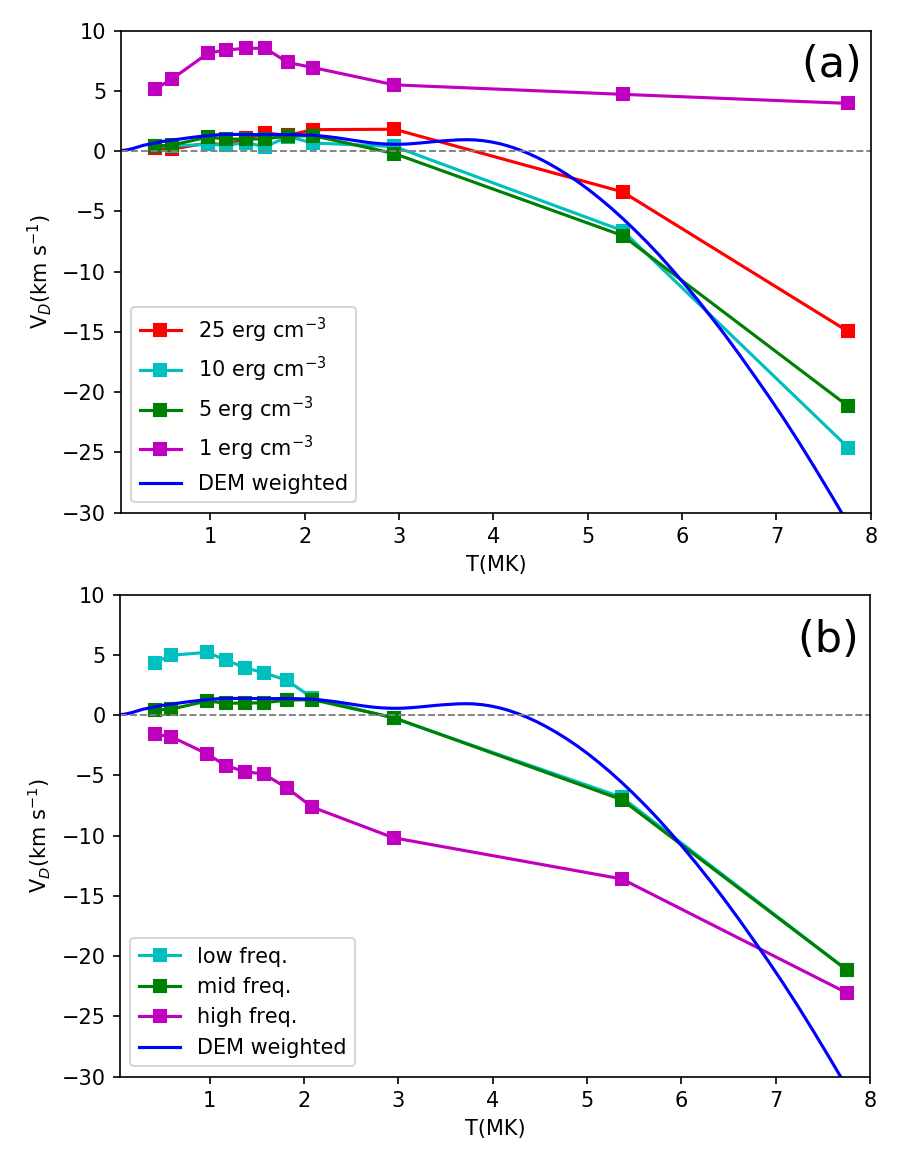}
       \caption{Doppler-shift velocities as a function of temperature for spectral lines integrated from the evolution of single nanoflares with different energies and frequency regimes. Panel a: Four nanoflares with different total volumetric energies as indicated in the panel legend (see the corresponding evolutions in Figure~\ref{fig:energy}). Panel b: Three integrations for a 5 erg cm$^{-3}$ nanoflare along 1000 s (corresponding to a high nanoflare frequency), 3000 s (middle frequency) and 5000 s (low frequency). The DEM weighted velocity integrated for the full 2DCAM-EBTEL model run is added for comparison (inverted with respect to Figure~\ref{fig:demv}).}
\label{fig:dop_vel_nano}
\end{figure*}

\begin{figure*}[ht!]
\centering
\hspace{0.cm}
\includegraphics[width=0.6\textwidth]{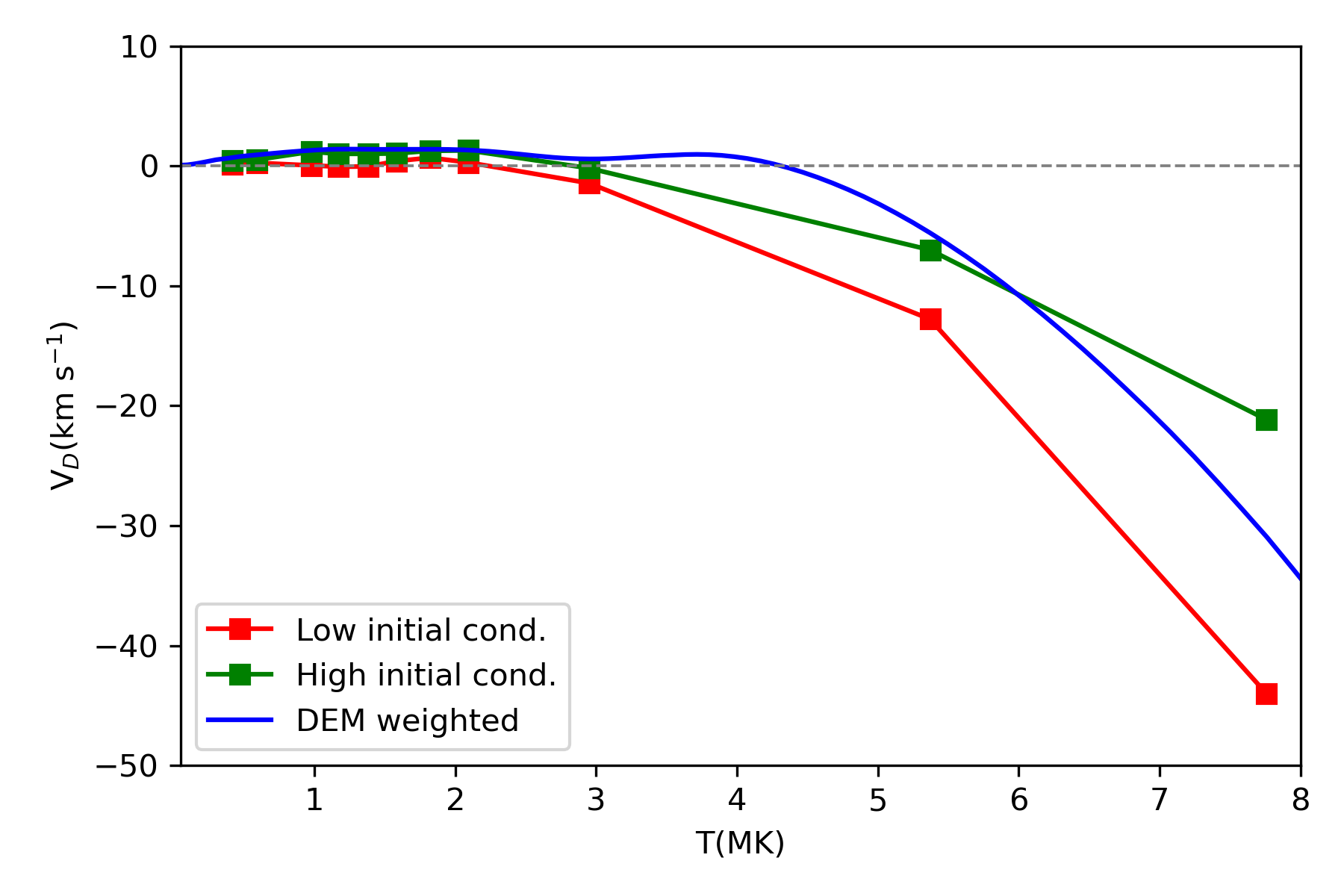}
       \caption{Doppler-shift velocities as a function of temperature for spectral lines integrated from the evolution of single nanoflares with different initial conditions. Low initial conditions: $T=$ 0.88 MK, $n=9\times10^{7}$ cm$^{-3}$, $v=0$. High initial conditions: $T=$ 1.8 MK, $n=2\times10^{9}$ cm$^{-3}$, $v=-10^{6}$ cm s$^{-1}$. The DEM weighted velocity integrated for the full 2DCAM-EBTEL model run is added for comparison (inverted with respect to Figure~\ref{fig:demv}).}
\label{fig:dop_vel_ini}
\end{figure*}

In this section we study the Doppler-shift velocities obtained with the spectral line analysis described in the last paragraphs of Section~\ref{sect:lines}, in particular, using Equations~\ref{eq:1stmom} and~\ref{eq:dopvel}. We apply the procedure to spectral lines integrated using five 30 s intervals from the 2DCAM-EBTEL output, as described in Section~\ref{sect:lines}. We call each of these sets of spectral lines ``samplings'' and we numerate them from 1 to 5. In Figure~\ref{fig:dop_vel_loop}a we plot, with different colors for each of the samplings (labeled S1 to S5), the Doppler-shift velocities of the modeled spectral lines as a function of temperature. The black squares and lines correspond to the mean values of the five samplings for each of the spectral lines. In the 4th and 5th columns of Table~\ref{table1} we list those mean values and the corresponding standard deviations as a reference for the relative dispersion of the Doppler shifts in each set of spectral lines. In Figure~\ref{fig:dop_vel_loop} (as in the successive Doppler-shift velocity plots in Figures~\ref{fig:dop_vel_nano} and~\ref{fig:dop_vel_ini}) the velocity sign convention is given by the definition from Equation~\ref{eq:dopvel}: positive velocities correspond to redshifts (motions away from the observer, i.e., downflows in on-disk observations) and negative velocities to blueshifts (i.e., upflows). Notice that this sign convention is opposite to the model-based convention of Figure~\ref{fig:demv}.

Inspection of Figure~\ref{fig:dop_vel_loop}a indicates that for the spectral lines with formation temperatures below 4 MK (the first nine lines of Table~\ref{table1}) there is a predominance of low velocity downflows (0.5 to 2.5 km s$^{-1}$), while for the two lines above 4 MK (FeXVII and FeXIX) Doppler shifts correspond to higher velocity upflows between 5 and 20 km s$^{-1}$. Downflows of a few km s$^{-1}$ have been reported from the analysis of spectral observations for temperatures below 1 MK \citep[see e.g.][]{winebarger2013}. Others have found downflows of approximately 10 km s$^{-1}$ at temperatures near 0.1 MK \citep{ghosh2019}, but these are likely to be associated with Type II spicules. These flows are faster than expected from nanoflare draining. Upflows between 1 MK and 1.5 MK have been reported by other authors \citep{warren2011b,tripathi2012a} with velocities up to 20 km s$^{-1}$. In contrast, with the parameters used in our model the shift from downflows to upflows occurs at temperatures around $\sim$ 3 MK \citep[see also,][]{tripathi2012b, peter1999}.

The behavior observed in Figure~\ref{fig:dop_vel_loop}a is summarized in the mean Doppler-shift values and standard deviations of the 4th and 5th columns of Table~\ref{table1}. Although there is an increasing dispersion of the Doppler-shift velocities with temperature, the relative variation, represented by the ratio of the standard deviation to the absolute value of the mean, varies between 1/3 and 1 with no apparent dependence on the temperature. 

As we began to discuss in Section~\ref{sect:lines}, since the spectral lines are obtained from the superposition of the contributions of the 49 strands of the model, we suggest that the variation of the Doppler-shift values obtained for the different spectral line samplings at different temperatures, reflect the variation of the detailed thermal and mass motion characteristics of the plasma in the strands during the sampled time intervals. This is confirmed by the analysis of individual nanoflare cases that follows.

In Figure~\ref{fig:dop_vel_loop}b we re-plot the mean Doppler-shift velocities from Figure~\ref{fig:dop_vel_loop}a (the black squares and lines identified as ``Mean'' in the legend) together with the Doppler-shifts obtained from the spectral lines integrated from the evolution of the single standard nanoflare described in Section~\ref{sect:nanoflares} (corresponding to the green curves in Figures~\ref{fig:energy} and~\ref{fig:init_cond}) and the 10$^4$ s strand evolution shown in Figure~\ref{fig:evolution}. As a reference, we also include the EM weighted mean velocity (the blue curve) reproduced from Figure~\ref{fig:demv}. The curve appears inverted with respect to Figure~\ref{fig:demv} due to the sign convention change mentioned previously. Note that the samplings mean follows very closely the EM weighted mean velocity for the lower temperatures, but for temperatures above 4 MK, although it shows the same tendency, the growth of the velocity absolute value is less pronounced. Notice from Figure~\ref{fig:dop_vel_loop}a though, that somewhat larger velocities are not unusual (see samplings 1, 4 and 5). The single strand and single nanoflare curves in Figure~\ref{fig:dop_vel_loop}b confirm this. The reason why the velocities obtained from the spectral line analysis differ from the expected EM weighted mean is mainly due to the averaging of several different strands going through different evolution stages, but also to the breadth of the spectral line contribution functions that yields the inclusion of emission from plasmas at a relatively broad range of temperatures. We suggest that this double averaging flattens the velocity curves at the higher temperatures. 

It is interesting to compare the single strand and single nanoflare curves of Figure~\ref{fig:dop_vel_loop}b. The similarity of these curves, despite the fact that the single strand evolution corresponds to four consecutive nanoflares whose evolutions differ from the single standard case, is remarkable. The only relevant difference is for the cooler spectral lines below 2 MK. This difference is likely due to the long cooling phase between the 3rd and the 4th nanoflares of the single strand evolution (see Figure~\ref{fig:evolution}), as suggested by the nanoflare frequency discussion below. The similarity of the single nanoflare curve with the mean of the model samplings and the single strand case confirms that the 5 erg cm$^{-3}$ single nanoflare with 3000 s evolution is a good choice for the typical nanoflare of the 2DCAM-EBTEL model.  

Let us now explore how the evolution of different types of nanoflares produced by the model affect the Doppler shifts of spectral lines. In Figure~\ref{fig:dop_vel_nano}a we plot the Doppler velocities obtained from spectral lines integrated over the 3000 s evolution of the four nanoflares with different total energies presented in Figure~\ref{fig:energy}. Notice that the curves for the 5, 10 and 25 erg cm$^{-3}$ nanoflares follow very closely the curve of the model DEM weighted mean velocity, in a similar way to the samplings of Figure~\ref{fig:dop_vel_loop} and their mean values. This suggests that those three energy levels represent fairly well the nanoflare evolution stages sampled from the strands of the model. 

The 1 erg cm$^{-3}$ case, on the other hand, is mostly flat around a downflow velocity of 5 km s$^{-1}$. This is consistent with the corresponding velocity evolution shown in Figure~\ref{fig:energy}, bottom panel (the purple curve), that indicates that the 1 erg cm$^{-3}$ case exhibits downflow velocities along all its evolution. The reason can be readily understood from the corresponding density evolution of the middle panel of Figure~\ref{fig:energy}. As we explained in Section~\ref{sect:nanoflares}, we chose to initiate the evolution of the four energy level nanoflares from initial conditions considered typical for the present model, which includes a 10 km s$^{-1}$downflow (draining after the previous event). Evaporation from the 1 erg cm$^{-3}$ energy release is not strong enough to produce an upflow against this headwind. The evolution of this case and the corresponding Doppler velocities obtained from the modeled spectral lines, suggest that under such initial conditions, low energy nanoflares do not make a substantial contribution to the multiple strand samplings of Figure~\ref{fig:dop_vel_loop}a. The reader might wonder what would be the case at different initial conditions, i.e., lower temperature and density. However, such initial conditions are not expected for low energy nanoflares. As it is nicely illustrated in the strand evolution of Figure~\ref{fig:evolution} (between the 3rd and the 4th nanoflares), after a long cooling process that leaves the plasma at low temperatures and densities it is generally expected that a higher than average energy nanoflare will occur, as the system has longer time to accumulate magnetic energy to be released. The above discussion suggests that despite being more frequent, the nanoflares at the lower end of the energy distribution contribute disproportionally less to the Doppler velocities obtained from the spectral lines modeled with 2DCAM-EBTEL. Below we analyze the effect of different initial conditions for the 5 erg cm$^{-3}$ nanoflare case. 

One last comment regarding Figure~\ref{fig:dop_vel_nano}a. Notice that for the high temperature end, the Doppler velocity of the 10 erg cm$^{-3}$ flare is higher in absolute value than for the 5 erg cm$^{-3}$ case. This is in principle expected, since from the evolution shown in Figure~\ref{fig:energy}, the velocity of the plasma at high temperature, for which the FeXIX line is more sensitive, is larger for the 10 erg cm$^{-3}$ case than for the 5 erg cm$^{-3}$ case. However, following the same argument the 25 erg cm$^{-3}$ nanoflare Doppler velocity should be even higher in absolute value, but it is in fact smaller. Inspecting again Figure~\ref{fig:energy} we see that the temperature of the plasma for the 25 erg cm$^{-3}$ nanoflare at the times of its maximum velocity is above 10 MK, and therefore farther from the formation temperature of the FeXIX line than the other two cases. In this case, the diminishing value of the contribution function of the line, for very high temperatures, produce a lesser effect on the global Doppler shift of the modeled spectral line. This example illustrates how the complexity of the combination of the plasma and the radiated emission properties produces different characteristics of the modeled spectral lines.   

In Figure~\ref{fig:dop_vel_nano}b we plot the Doppler velocities of the spectral lines constructed by integrating the standard 5 erg cm$^{-3}$ nanoflare simulation over 1000 s (high frequency), 3000 s (mid frequency) and 5000 s (low frequency) as described in Section~\ref{sect:nanoflares}. The respective colors for the three cases are identified in the Figure legend. The DEM weighted mean velocity of the 2DCAM-EBTEL model is once again included for comparison. The main evident features of these plots are the clearly different behavior of the low temperature Doppler velocities (below 2 MK) for the three cases, and the similarity of the low and middle frequency cases for temperatures above 2 MK. It can be also seen that despite the clear difference between the high frequency case and the other two for all spectral lines, they tend to be more similar at the high temperature end of the plot. The reader is reminded that these cases represent one impact of nanoflare frequency - the amount of cooling that can occur before reheating. The other impact is initial conditions, which we discuss shortly.

The features just described are easy to understand from the evolution of the 5 erg cm$^{-3}$ nanoflare shown in Figures~\ref{fig:energy} and~\ref{fig:init_cond} (the green curves). The spectral line integration for the high frequency case includes only the first 1000 s of the evolution, when the temperature is relatively high and the velocity corresponds exclusively to upflows, as it is reflected in the Doppler velocities obtained. As the plasma cools down and the contribution to the cooler lines increases, the velocity decreases and smaller Doppler velocities are obtained for those lines. The integration for the mid and low frequency cases coincide up to the first 3000 s of the evolution, when temperatures are between 2.5 and 7 MK and velocities correspond to both upflows (for the higher temperatures) and downflows (for the lower temperatures). Since they coincide in the high and middle temperature range they also have very similar Doppler velocities in that range in Figure~\ref{fig:dop_vel_nano}b. In the case of the higher temperature, where the integration also coincides with the high frequency case, the Doppler velocities for the FeXIX line ($T \approx$ 7.8 MK) tend to be more similar. Finally, for the last 2000 s of the evolution of the low frequency case (3000 s $<t<$ 5000 s, not included in Figures~\ref{fig:energy} and~\ref{fig:init_cond}), we have the lowest temperatures and highest velocity downflows that contribute to the difference between the mid and low frequency Doppler velocities for temperatures below 2 MK in Figure~\ref{fig:dop_vel_nano}b. Interestingly, these higher velocity downflows (4-5 km s$^{-1}$) observed at the lower temperatures for the low-frequency nanoflare case are similar to the single strand integration case shown in Figure~\ref{fig:dop_vel_loop}b. The reason for this similarity is the presence of a long cooling phase between the 3rd and 4th nanoflares of the single strand evolution (see Figure~\ref{fig:evolution}), which resembles the 5000 s evolution of the low-frequency nanoflare case, in particular, the last 2000 s of evolution whose contribution mainly affect the cooler spectral lines. 

Notice that the $\sim$5 km s$^{-1}$ downflows at lower temperatures observed in the single strand example and the low frequency nanoflare do not appear to have a noticeable contribution in the samplings of Figure~\ref{fig:dop_vel_loop}a, where we see Doppler velocities below 1 or 2 km s$^{-1}$ (see also the mean values in the fourth column of Table\ref{table1}). This is explained by the relatively low abundance of low frequency nanoflares (7\%) in the 2DCAM-EBTEL model, for the 100 Mm loop case studied here, as mentioned in Section~\ref{sect:nanoflares}. Although we chose the single strand of Figure~\ref{fig:evolution} as a good example for presenting nanoflares with the three different nanoflare frequency scenarios, the long cooling phase between the third and the fourth nanoflares is not typical of the studied model.

We now analyze the effect on the Doppler-shifts of nanoflare evolutions with different initial conditions, as discussed in Section~\ref{sect:nanoflares}. In Figure~\ref{fig:dop_vel_ini} we plot the Doppler-shift velocities obtained from the integration of spectral lines over the nanoflare evolutions presented in Figure~\ref{fig:init_cond}. We use the same color coding for the curves in both figures. The most notable difference between the Doppler-shift curves for the low and high initial condition nanoflares observed in Figure~\ref{fig:dop_vel_ini}, is their departure at the highest temperatures, being the Doppler velocity for the FeXIX 7.6 MK line a factor two larger for the low initial conditions case. This is readily understood by inspecting the evolution curves in Figure~\ref{fig:init_cond}. As usual, the higher upflow velocities occur at the higher temperatures. Notice however that the maximum velocity for the low initial conditions case is approximately a factor 20 larger than for the high initial conditions case. We remind the reader that both nanoflares are the same in terms of total energy and heating injection rate (5 erg cm$^{-3}$ and a triangular function of 200 s duration).   

The key difference between the observed evolutions is the density at the start of the nanoflare. We can assume that comparatively little energy is radiated from the strand during the 200 s of the nanoflare. The thermal energy content of the strand therefore increases by an amount equal to total nanoflare energy release. Thermal energy density is proportional to pressure, but since density is relatively constant due to limited time for evaporation to operate, the change is primarily a temperature increase. The amount of increase is inversely proportional to the initial density. Smaller densities that are characteristic of low frequency nanoflares result in hotter peak temperatures. Note, however, that thermal conduction cooling may limit the peak temperature. The temperature will not rise above the point where the cooling equals the instantaneous heating rate of the nanoflare. Thermal conduction losses are a strong function of temperature, varying as $T^{7/2}$, so they can quickly become significant.

Notice that despite the approximate factor 20 difference between the velocities observed at the impulsive ``hot'' phase at the beginning of the evolutions (see the bottom panel of Figure~\ref{fig:init_cond}), the difference between the obtained Doppler-shift velocities at the higher temperatures is only about a factor 2. Also, while the maximum velocity reached by the plasma in the ``low'' case is close to 1000 km s$^{-1}$, the Doppler shift velocity of the hotter line has an absolute value of 44 km s$^{-1}$. As discussed before, this is caused by the smearing effect of the line contribution function breadth and the rapidly evolving temperature of the plasma during the nanoflare impulsive phase.

The analysis of this Section shows that there is a complex interplay between the spectral line formation temperatures, the width of the line contribution functions, the different properties of the plasma along the nanoflares evolution and, in the case of the 2DCAM-EBTEL model samplings, the sum of the contribution of different strands at different stages of their evolutions. As explained in Section~\ref{sect:lines}, we construct the spectral lines computing separately the TR and coronal contributions from the respective parameters provided by the EBTEL model. As discussed in Section~\ref{sect:demv} and suggested by the results shown in Figure~\ref{fig:demv}, the relative weight of the TR and coronal contributions to the spectral lines vary along the studied temperature range, being one of the causes for the ample Doppler-shift velocity variations observed. In the sixth column of Table \ref{table1} we show the mean relative weight of the TR and coronal intensity contributions, averaged over the 2DCAM-EBTEL samplings, for the different ion temperatures. There, it is very clear the overwhelming dominance of the TR contribution for the cooler lines by a factor of more than 400, while the opposite is true for the hotter lines, where the coronal intensity is more than 30 times stronger. The change occurs between the FeXV 2.09 MK line and the CaXIV 2.95 MK line, around where both contributions have comparable strengths. This roughly coincides with the intermediate zone of the EM weighted mean velocity plot of Figure \ref{fig:demv}, where both contributions are comparable. It is necessary to emphasize that the relative weight of the TR and coronal contributions is due to a combination of the width of the line contribution functions and the relative density of the TR and the corona at the different corresponding temperatures. We note that the relative brightness of the TR and corona depends on the length of the strand in addition to the properties of the nanoflare \citep{schonfeld2020}.

\subsection{Line asymmetry}
\label{sect:asym}

Another characteristic usually computed in plasma diagnostics of observed spectral lines is the relative weight of the red and blue wings of the lines. Here, we define an asymmetry parameter of the line in the following way. We divide the line profile in two wings at both sides of the maximum intensity, the left, or blue wing, and the right, or red wing. Then, we compute the difference between the integrals of the red and the blue wings and divide it by the full integral of the line. With this definition the asymmetry parameter has the same sign as the skewness of the profile. If the result is positive there is a predominance of the red wing over the blue wing and if the opposite is true the blue wing dominates. The dependence of this parameter with the temperature is another testable prediction of the nanoflare model.

For lines from FeVIII to CaXIV we compute the asymmetry parameter using a wavelength range of width 0.3 Å, centered in the line formation wavelength, corresponding to velocity widths between 160 and 230 km s$^{-1}$ depending on the line. For the broader hotter lines, FeXVII and FeXIX, the used velocity range widths correspond to approximately 600 and 500 km s$^{-1}$, respectively.

\begin{figure*}[ht!]
\centering
\hspace{0.cm}
\includegraphics[width=0.6\textwidth]{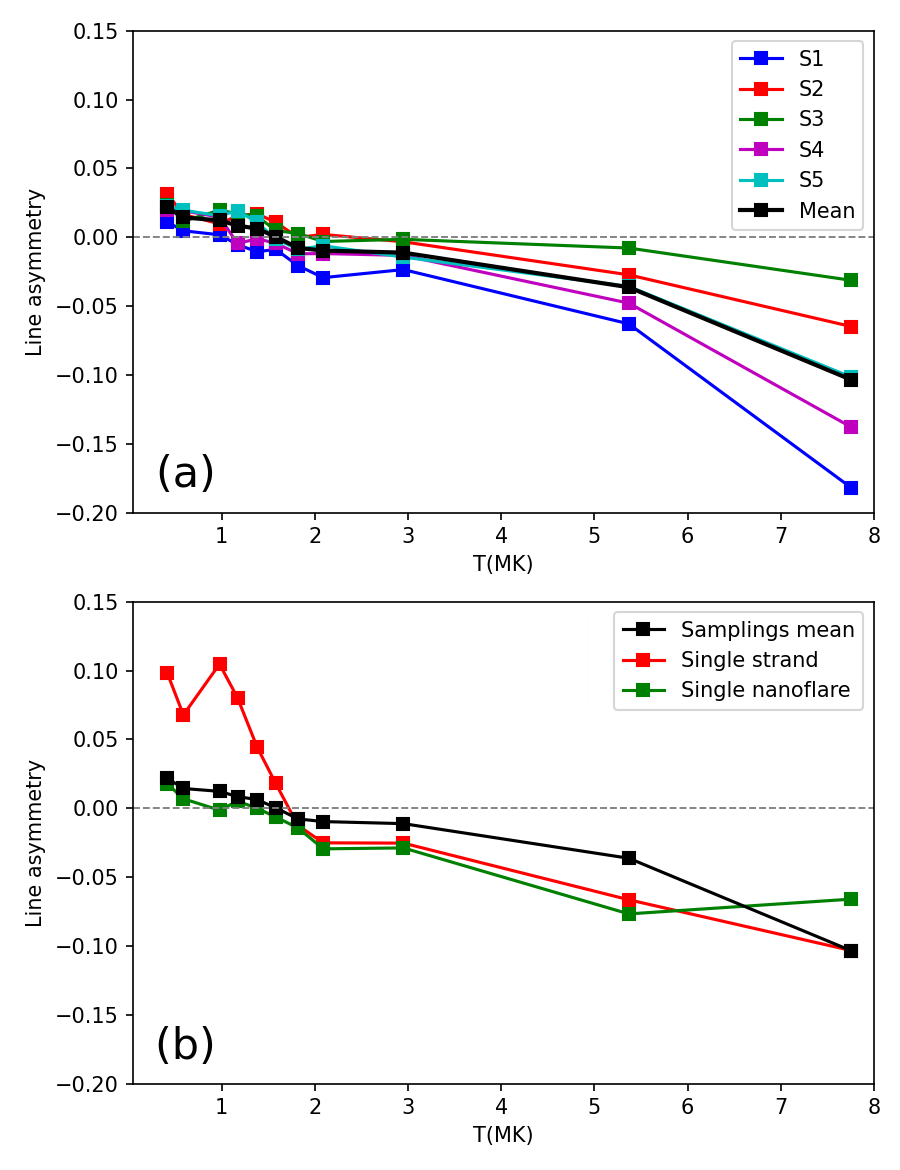}
       \caption{Asymmetry parameter as a function of temperature for spectral lines integrated from the evolution of single and multiple modeled strands. Panel a: 30 s samplings from the 2DCAM-EBTEL model. The black squares and lines correspond to the mean values. Panel b: Comparison of the mean of panel a (the black curve) with the asymmetry parameters of lines integrated from the 10$^{4}$ s single strand evolution shown in Figure~\ref{fig:evolution} (red curve) and the single 5 erg cm$^{-3}$ standard nanoflare evolution of Figure~\ref{fig:energy} (green curve).}
\label{fig:asym_loop}
\end{figure*}

In Figure~\ref{fig:asym_loop}a we plot, as a function of temperature, the line asymmetry parameter computed for the five 2DCAM-EBTEL line samplings described in Section~\ref{sect:lines}. As in the Doppler-shift velocity plots of Figures~\ref{fig:dop_vel_loop}, the black squares and lines correspond to the mean values of the samplings. Inspection of the figure indicates that cooler spectral lines have small positive asymmetry parameters, indicating a moderate redshift dominance for those lines. In the other extreme, at higher temperatures, the sampled spectral lines have larger and more spread asymmetry parameter values. In the 7th column of Table~\ref{table1} we list the mean of the asymmetry parameter values of the five samplings for the different modeled spectral lines. The dependence of the asymmetry parameter on the temperature is generally similar to the Doppler-shift velocities plotted in Figure~\ref{fig:dop_vel_loop}a, where the lower temperatures associate with low velocity redshifts and the higher temperatures with higher velocity blueshifts.
As seen in Figure~\ref{fig:asym_loop}a and the corresponding column of Table~\ref{table1}, the mean value change from redshifts to blueshifts occurs approximately at $T = 1.58$ MK, corresponding to the FeXIII ion. Notice in Figure~\ref{fig:dop_vel_loop}a however, that this change does not occur at the same temperature for all samplings; for example, while for S1 the change occurs at approximately 1 MK (between FeX and FeXI), for S3 it occurs at around 3 MK (CaXIV). This indicates, once again, that some of the spectral line characteristics are determined by the particular details of the evolution of the strands that contribute to the emission during the sampling. This is specially the case at intermediate temperatures where the TR and coronal contributions are comparable. 

For comparison, similarly to Figure~\ref{fig:dop_vel_loop}b, in Figure~\ref{fig:asym_loop}b we replot the mean values of the line asymmetry parameters of the samplings (the black line), together with the single nanoflare and the single strand cases. The three curves follow roughly the same behavior, the exception being the asymmetry parameter of the single strand case at the lower temperatures, where it takes higher positive values. The origin of this is likely the high velocity downflows at low temperatures produced during the strand evolution between the 3rd and the 4th nanoflares (see Figure~\ref{fig:evolution}), which produce an enhancement of the redshift contribution at the lower temperature spectral lines. This is similar to what is observed in the Doppler-shift velocity plot of Figure~\ref{fig:dop_vel_loop}b.  

Regarding the curve for the standard nanoflare (the green curve in Figure~\ref{fig:asym_loop}b), notice the inversion of the growing of the absolute value of the asymmetry parameter at the highest temperature, contrary to what is observed for the rest of the curves. This is likely due to the particular evolution of the standard 5 erg cm$^{-3}$ nanoflare, as we see below. 

In Figure~\ref{fig:asym_nano} we compare the asymmetry parameter for the test nanoflares with different energies (panel a) and nanoflare frequencies (panel b). 
In Figure~\ref{fig:asym_nano}a the 5, 10 and 25 erg cm$^{-3}$ cases show a behavior that is generally similar to the sampling cases of Figure~\ref{fig:asym_loop}a, with small positive asymmetry parameters for the lowest temperatures, near zero for intermediate temperatures and increasingly negative asymmetry parameters of the higher temperatures. As it is evident comparing with Figure~\ref{fig:asym_loop}b, in the intermediate range between 1 and 2 MK, the 5 erg cm$^{-3}$ case is the one that better resembles the 2DCAM-EBTEL sampling average. Similarly to what is observed in the Doppler-shift velocities plot of Figure~\ref{fig:dop_vel_nano}, the 1 erg cm$^{-3}$ case behavior differs substantially from the other three cases, showing large asymmetry parameters for the lowest temperatures and around 0 values for the higher temperatures starting at 1.5 MK. As we discussed in Section~\ref{sect:doppler}, this behavior is produced by the particular evolution of the 1 erg cm$^{-3}$ nanoflare. Notice, in particular, that the high asymmetry parameters of the lower temperatures is produced by the strong downflows at low temperatures at the end of its 3000 s evolution (see the magenta curve in the bottom panel of Figure~\ref{fig:energy}). 

\begin{figure*}[ht!]
\centering
\hspace{0.cm}
\includegraphics[width=0.6\textwidth]{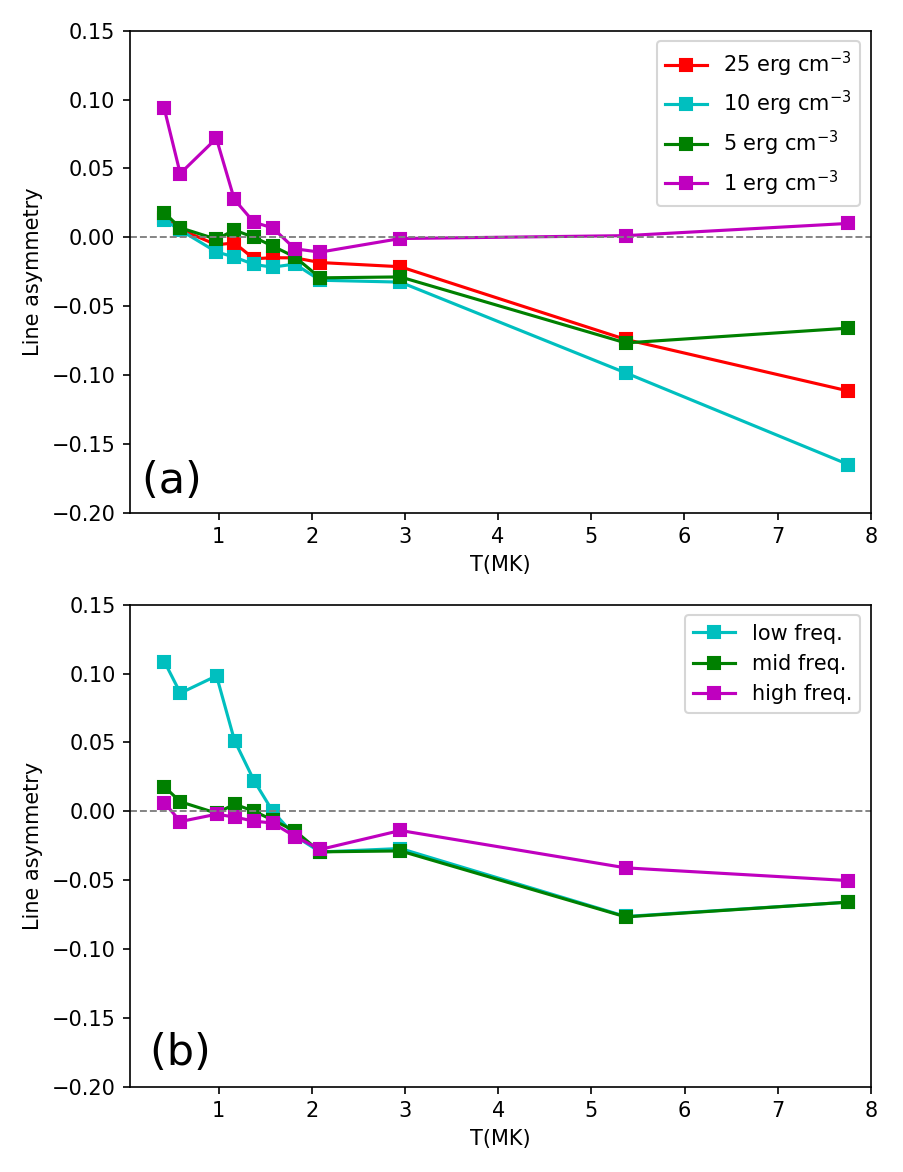}
       \caption{Asymmetry parameter as a function of temperature for spectral lines integrated from the evolution of single nanoflares with different energies and frequency regimes. Panel a: Four nanoflares with different total volumetric energies as indicated in the panel legend (see the corresponding evolutions in Figure~\ref{fig:energy}). Panel b: Three integrations for a 5 erg cm$^{-3}$ nanoflare along 1000 s (corresponding to a high nanoflare frequency), 3000 s (middle frequency) and 5000 s (low frequency).}
\label{fig:asym_nano}
\end{figure*}

In Figure~\ref{fig:asym_nano}b the three nanoflare frequency cases studied show that for low temperatures the middle and high frequency cases have similarly small asymmetry parameters, although in some cases with different signs, as for the SiVII line at 0.59 MK and FeXI at 1.17 MK, in which the middle-frequency asymmetry parameter is positive and the high frequency case is negative. Notice that this is different from the behavior observed in the Doppler-shift velocities of Figure\ref{fig:dop_vel_nano}b, in which the high and middle frequency cases do not coincide anywhere along the analyzed temperature range. 

In the low temperature range (up to 1.38 MK) the low frequency nanoflare case has much larger asymmetry parameters. As in the cases of the single strand evolution and the 1 erg cm$^{-3}$ nanoflare (see Figures~\ref{fig:asym_loop}b and~\ref{fig:asym_nano}a), this behavior is due to the high downflow velocities reached at very low temperatures at the final stages of the 5000 s low frequency nanoflare evolution. The absence of this extreme behavior in the asymmetry parameters of the 2DCAM-EBTEL samplings of Figure~\ref{fig:asym_loop}a is, once again, due to the fact that low frequency cases amount to just 7\% of the nanoflares in the 2DCAM-EBTEL 100 Mm loop model. However, the presence of positive asymmetry parameters at temperatures up to 2 MK, observed in some of the samplings, may be due to a non-negligible contribution of nanoflares presenting relatively strong downflows at low temperatures.

For the temperatures in the intermediate range: 1.58, 1.82 and 2.09 MK (corresponding to FeXIII, FeXIV and FeXV, respectively), the asymmetry parameters of the three nanoflare frequency cases presented in Figure~\ref{fig:asym_nano}b approximately coincide, indicating increasing upflow dominance. Finally, for the higher temperatures: 2.95, 5.37 and 7.76 MK (corresponding to CaXIV, FeXVII and FeXIX) the middle and low frequency cases exactly coincide, while the high frequency case presents a slightly smaller upflow dominated asymmetry. Notice that the reverse tendency of the middle frequency nanoflare mentioned previously in the description of Figure~\ref{fig:asym_loop}b is also observed in the low frequency case (although not in the high frequency nanoflare!), suggesting that that behavior is due to the temperature and velocity contributions at the time ranges when the integration of the low and middle frequency cases coincide, that is to say, between $t=$ 1000 s and $t=$ 3000 s along the 5 erg cm$^{-3}$ nanoflare evolution.

\begin{figure*}[ht!]
\centering
\hspace{0.cm}
\includegraphics[width=0.6\textwidth]{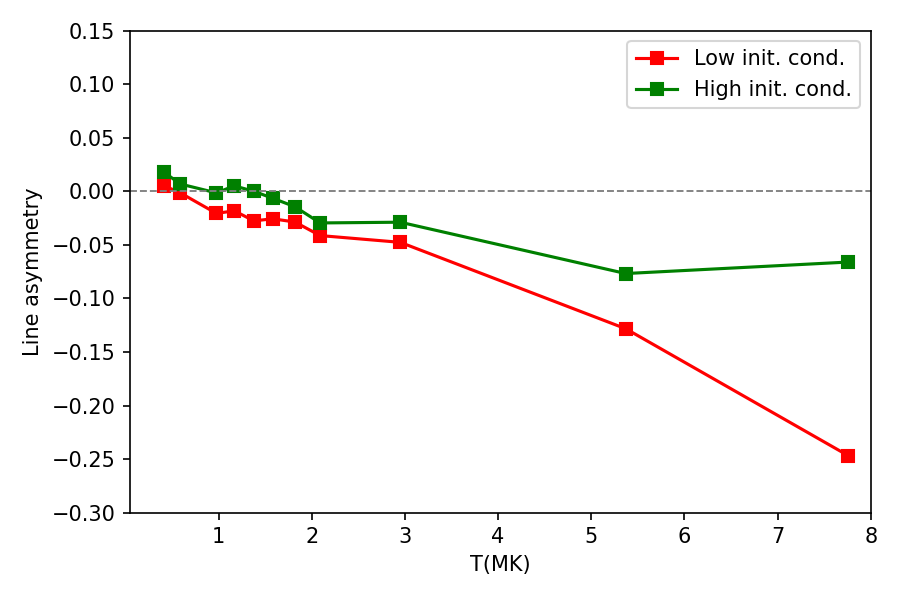}
       \caption{Asymmetry parameter as a function of temperature for spectral lines integrated from the evolution of single nanoflares with different initial conditions. Low initial conditions: $T=$ 0.88 MK, $n=9\times10^{7}$ cm$^{-3}$, $v=0$. High initial conditions: $T=$ 1.8 MK, $n=2\times10^{9}$ cm$^{-3}$, $v=-10^{6}$ cm s$^{-1}$.}
\label{fig:asym_init_cond}
\end{figure*}

For completeness, in Figure~\ref{fig:asym_init_cond} we also include the comparison of the asymmetry parameters for the low and high initial condition nanoflares described in Section~\ref{sect:nanoflares}. The asymmetry parameters of the low initial condition nanoflare lines show very little redwing dominance only for the lowest temperature, and increasing bluewing dominance towards higher temperatures, similarly to the 10 erg cm$^{-3}$ nanoflare case shown in Figure~\ref{fig:asym_nano}a. This similarity is understood by inspecting the respective evolution curves, in particular, the temperature, in Figures~\ref{fig:energy} and~\ref{fig:init_cond}. Notice however that while the maximum velocity at high temperatures is, in the case of the low initial conditions case, around 1000 km s$^{-1}$ and the corresponding asymmetry parameter is 0.25, for the 10 erg cm$^{-3}$ nanoflare its maximum velocity is of the order of 100 km s$^{-1}$ with a corresponding asymmetry parameter of approximately 0.15. As in the case of the Doppler-shift velocities, discussed in Section~\ref{sect:doppler}, this lack of proportion between the actual velocities and the measured line parameters is likely due to the smearing of wide thermal contribution functions and the integration of varying physical quantities along the nanoflare evolution.

Once again, the study of single nanoflares confirms that the spectral line properties obtained from the 2DCAM-EBTEL samplings are consistent with the simultaneous contributions of strands with a variety of evolutions. The analysis helps us understand which features of the nanoflare evolutions define the observed line properties.

\subsection{Non-thermal velocity}

As discussed at the end of Section~\ref{sect:lines}, the broadening of spectral lines can be related to the presence of non-thermal processes such as oscilations, turbulence and flows \citep[see e.g.,][and references therein]{patsourakos2006}. Here, we analyze the width of the modeled spectral lines in order to study how nanoflare flows produce such broadenings. A measure of the line broadening associated with non-thermal processes is given by the $\xi$ parameter from Equation~\ref{eq:nontherm},

\begin{equation}
\xi = \left(\frac{2c^2}{\lambda^2}M_2-\frac{2kT}{m_i}\right)^{\frac{1}{2}}.
\end{equation} 

\noindent This parameter has velocity units and is usually called non-thermal velocity, in contrast with the thermal velocity, $V_0 = (2kT/m_i)^{1/2}$. To compute $\xi$ from the above expression, we obtain the second moment of the line, $M_2$ (see Equation~\ref{eq:2ndmom}), we use $\lambda=\lambda_0$, the line central wavelength for the ion at rest, and we consider $T=T_f$, the line formation temperature (i.e., the temperature at which the line contribution function has its maximum). Notice that the broadening actually occurs, and $\xi$ is defined, only if the second moment of the line profile, $M_2$, is:
\begin{equation}
M_2 \ge \frac{kT_f\lambda_0^2}{m_ic^2}.
\label{eq:non-therm_cond}
\end{equation} 

\noindent Since the line profiles are constructed from the time integrated contributions of instantaneously defined lines of half-width $w_j(t)$, as indicated by Equation~\ref{eq:width}, if the temperature of the evolving plasma never reaches $T_f$, the half-width of the resulting profile, $M_2$, could be, under certain circumstances, smaller than the right hand side of Equation~\ref{eq:non-therm_cond}. In that case, we consider that no broadening is present and we set $\xi=0$. As we show below, this is actually the case of some of the studied synthetic hot lines. It is worth explaining what we mean above by ``certain cirumstances''. Even if the plasma does not reach $T_f$, a broadening produced by variable Doppler shifts and the summed contribution of different strands, could still produce a line profile width that accomplishes the relation given in Equation~\ref{eq:non-therm_cond}.

\begin{figure*}[ht!]
\centering
\hspace{0.cm}
\includegraphics[width=0.6\textwidth]{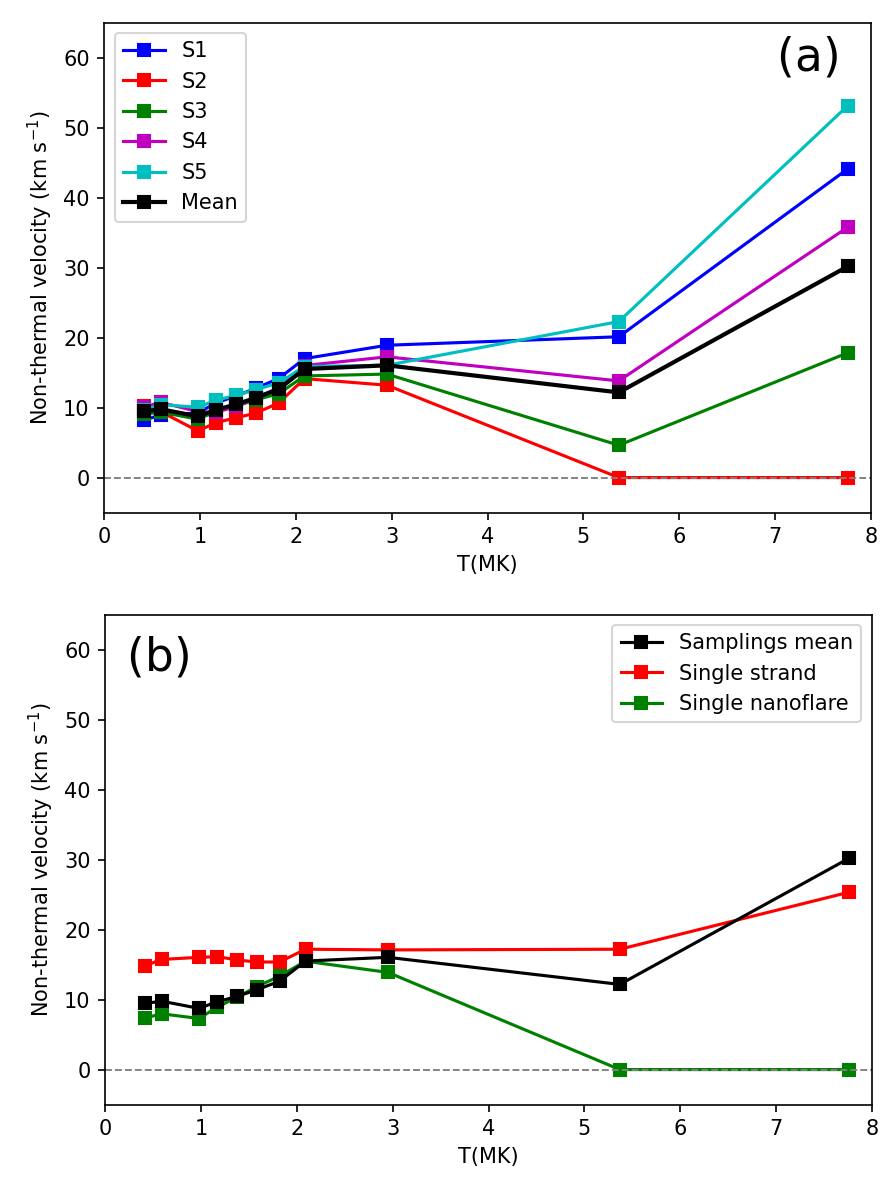}
       \caption{Non-thermal velocity as a function of temperature for spectral lines integrated from the evolution of single and multiple modeled strands. Panel a: 30 s samplings from the 2DCAM-EBTEL model. The black squares and lines correspond to the mean values. Panel b: Comparison of the mean of panel a (the black curve) with the non-thermal velocities of lines integrated from the 10$^{4}$ s single strand evolution shown in Figure~\ref{fig:evolution} (red curve) and the single 5 erg cm$^{-3}$ standard nanoflare evolution of Figure~\ref{fig:energy} (green curve).}
\label{fig:nontherm_loop}
\end{figure*}

Following the same scheme as in figures of previous sections, in Figure~\ref{fig:nontherm_loop}, panel a, we plot, as a function of temperature and with different colors, the non-thermal velocity parameters computed from the synthetic lines obtained from the 2DCAM-EBTEL samplings described in Section~\ref{sect:lines}. The black squares and lines correspond to the mean of the sampling values. We reproduce these mean values in the 9th column of Table~\ref{table1}. For comparison, in the 8th column we present the thermal velocity components of the corresponding lines computed from Equation~\ref{eq:therm}. As indicated before, for that computation we use the line formation temperature, $T_f$.

The most notable feature of the plots in Figure~\ref{fig:nontherm_loop}a is the little dispersion of the different samplings for temperatures below 3 MK and the wide spread of the curves for the hotter lines, corresponding to a full variation of non-thermal velocities of around 50 km s$^{-1}$ for the FeXIX line at $T=$ 7.76 MK. This is somehow similar to what is observed in the case of the Doppler shift velocities and the asymmetry parameters described in the previous sections - the higher the temperature, the more pronounced the variation of the parameters. Notice also that in the case of S2, for the FeXVII (5.37 MK) and FeXIX (7.76 MK) lines, we set $\xi=0$, since, as we discussed before, it is a particular case in which the width of the lines do not accomplish the relation given in Equation~\ref{eq:non-therm_cond}. It is also interesting to note that that for the hotter lines, the span of variation of the non-thermal velocities is uniformly filled with the different samplings, from the zero values of S2 to the maximum values of S5. 

The fact that the non-thermal velocities present, as the Doppler velocities and asymmetry parameters, very little change for cold lines and a wider variation for hot lines, indicates that in the 2DCAM-EBTEL model the plasma evolution is much more variable at high temperatures than at low temperatures. As before, this is explained by the analysis of the variety of individual nanoflares that make up the studied system.   

In Figure~\ref{fig:nontherm_loop}b we plot, for comparison, the non-thermal velocity against $T$ for the samplings mean, the standard 5 erg cm$^{-3}$ nanoflare and the single strand evolution of Figure~\ref{fig:evolution}. Notice that the standard single nanoflare is another case for which the two hottest lines have $\xi=0$. For the coolest temperatures though, it coincides very closely with the samplings mean. The evolution of the single strand with several nanoflares, on the other hand, follows more closely the samplings mean for the hot lines and departs from the mean, and hence from the samplings of panel b, at the coolest temperatures. This behavior is due, as in the cases of the Doppler velocities and asymmetry parameters studied in the previous subsections, to the long cooling phase between the third and the fourth nanoflares in the evolution shown in Figure~\ref{fig:evolution}. This is confirmed by the analysis of the low frequency nanoflare discussed below.

\begin{figure*}[ht!]
\centering
\hspace{0.cm}
\includegraphics[width=0.6\textwidth]{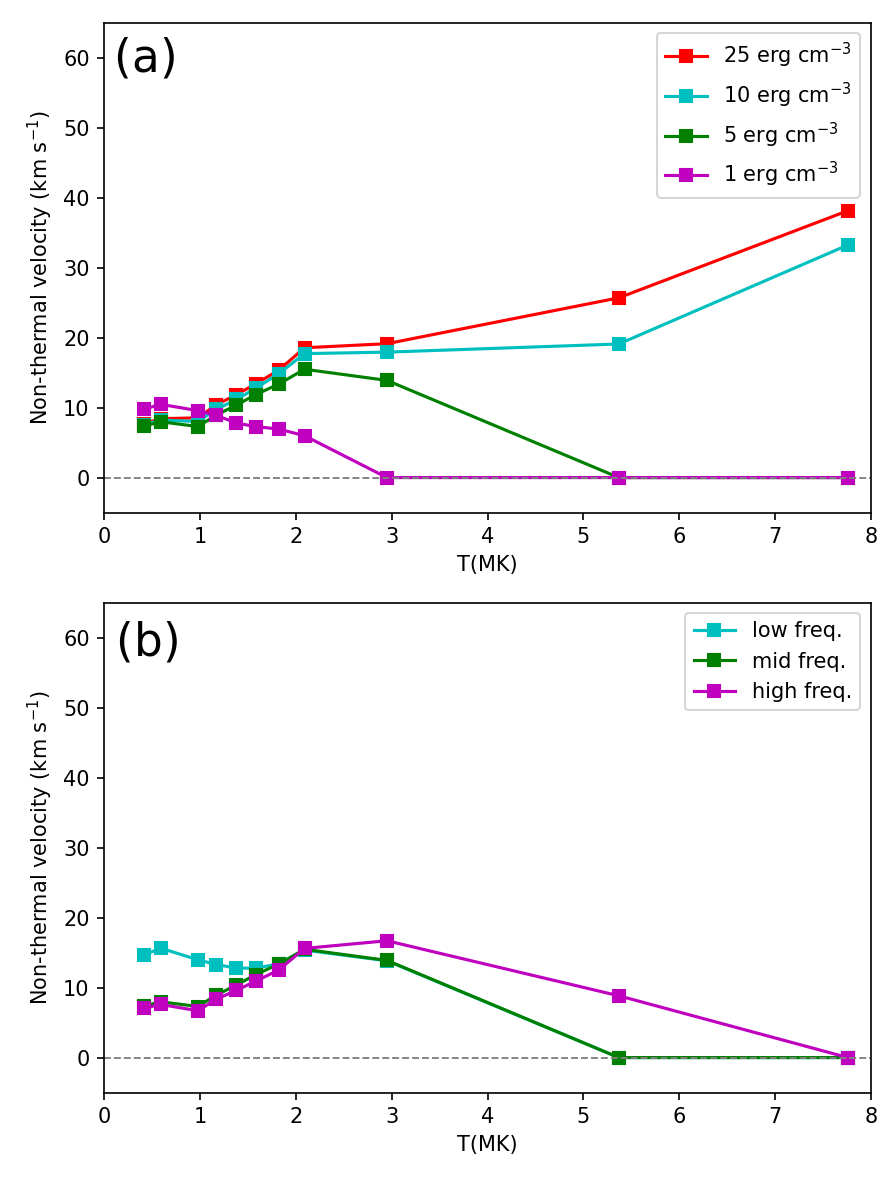}
       \caption{Non-thermal velocity as a function of temperature for spectral lines integrated from the evolution of single nanoflares with different energies and frequency regimes. Panel a: Four nanoflares with different total volumetric energies as indicated in the panel legend (see the corresponding evolutions in Figure~\ref{fig:energy}). Panel b: Three integrations for a 5 erg cm$^{-3}$ nanoflare along 1000 s (corresponding to a high nanoflare frequency), 3000 s (middle frequency) and 5000 s (low frequency).}
\label{fig:nontherm_nano}
\end{figure*}

In Figure~\ref{fig:nontherm_nano}a we plot the non-thermal velocities of individual nanoflares with different total volumetric energies. It is interesting to see that up to 1 MK the non-thermal velocities of the 1 erg cm$^{-3}$ case are similar to the mean values of the samplings of Figure~\ref{fig:nontherm_loop}a, while it rapidly diminishes for higher temperatures, resulting in $\xi=0$ for the three hotter lines. The similarity with the 2DCAM-EBTEL samplings at the lower temperatures suggest a non-negligible effect of low energy nanoflares on the model. 

For the other three nanoflare energies the non-thermal velocities are more similar to each other, and slightly smaller than the 1 erg cm$^{-3}$ case, but grow monotonously for temperatures between 1 and 2 MK, following the same tendency observed in all the samplings of Figure~\ref{fig:nontherm_loop}a. Above the CaXIV line temperature (2.95 MK) the 5 erg cm$^{-3}$ case departs from the other two taking $\xi = 0$ for the higher temperatures as we have seen in Figure~\ref{fig:nontherm_loop}b. The non-thermal velocities for the 10 and 25 erg cm$^{-3}$ cases continue to grow up to more than 30 km s$^{-1}$ for the FeXIX line (7.76 MK), which is consistent with the range observed for the 2DCAM-EBTEL samplings of Figure~\ref{fig:nontherm_loop}. 

In Figure~\ref{fig:nontherm_nano}b we plot the non-thermal velocity for the three different nanoflare frequency cases. We see that for the lower temperatures (up to $\approx$ 2 MK) the non-thermal velocity of the high frequency nanoflare coincides with the standard middle frequency case, with values varying from less than 10 to $\approx$ 15 km s$^{-1}$. For the same temperature range, the low frequency case departs from the other two at the lowest temperatures, with higher non-thermal velocity values, and tends to almost coincide with them at the temperatures of the FeXIV (1.82 MK) and FeXV (2.09 MK) lines. At CaXIV and FeXVII temperatures (2.95 and 5.37 MK, respectively), the three cases depart again but, in this case, it is the middle and low frequency nanoflares that coincide. At 7.76 MK (FeXIX) the three nanoflare frequency cases have a non-thermal velocity of 0. 

As before, the described behavior can be understood by comparing the integrated evolution of the high, middle and low frequency nanoflares (1000, 3000 and 5000 s evolutions, respectively). The non-thermal velocities computed at 1.82, 2.09 and 7.76 MK, where the three cases almost coincide, are produced within the first 1000 s of evolution, where the three line integrations also coincide. This is also true for the similarity of the high and middle frequency cases at the lowest temperatures, since that is the only time at which both integrations coincide at low temperatures. On the other hand, at these low temperatures, the low frequency case is dominated by the late cooling phase evolution (at $t>$ 3000 s) that presents the highest downflow velocities at low temperatures, a stage that only the low frequency case integration covers. The similarity of the low and middle frequency cases at intermediate temperatures of 2.95 and 5.37 MK corresponds to the evolution between 1000 and 3000 s where their integrations coincide. All these results are consistent with the Doppler shift and asymmetry parameter analysis of the previous sections.      

\begin{figure*}[ht!]
\centering
\hspace{0.cm}
\includegraphics[width=0.6\textwidth]{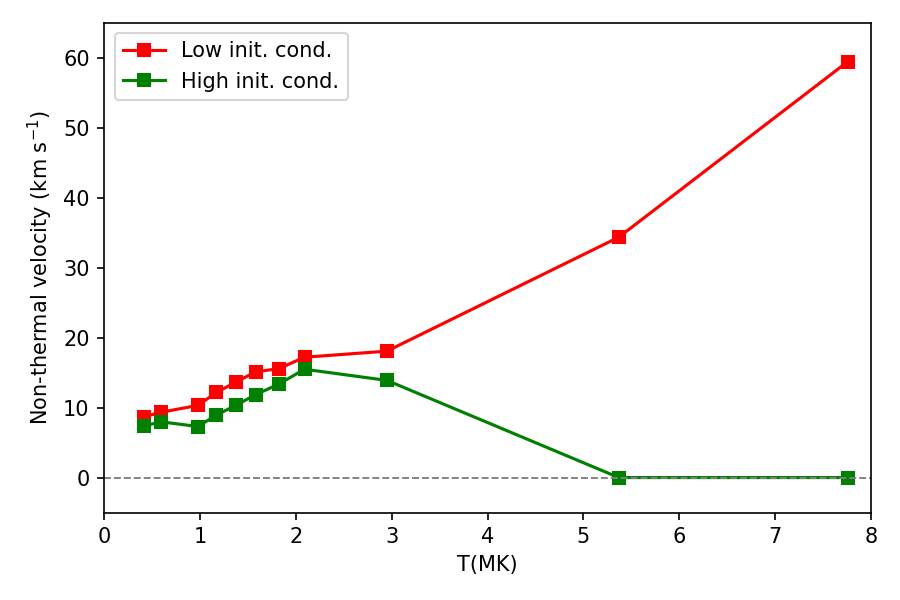}
       \caption{Non-thermal velocity as a function of temperature for spectral lines integrated from the evolution of single nanoflares with different initial conditions. Low initial conditions: $T=$ 0.88 MK, $n=9\times10^{7}$ cm$^{-3}$, $v=0$. High initial conditions: $T=$ 1.8 MK, $n=2\times10^{9}$ cm$^{-3}$, $v=-10^{6}$ cm s$^{-1}$.}
\label{fig:nontherm_init_cond}
\end{figure*}

Finally, in Figure~\ref{fig:nontherm_init_cond} we compare the non-thermal velocities for nanoflares with low and high initial conditions. We see that for temperatures up to $\approx$2 MK both cases show a similar behavior, except for the fact that the non-thermal velocities of the low initial conditions case are slightly higher. This is expected according to the evolutions shown in Figure~\ref{fig:init_cond}, considering that at the times when the temperature and velocity of the two cases differ, both plasma properties are systematically larger for the low initial conditions nanoflare than for the high initial conditions case. The difference is dramatic at temperatures above 3 MK, where the temperature and velocity variation of the  
low initial conditions nanoflare are substantially larger. This translates into a broad difference of non-thermal velocities at the highest temperatures, where the low initial conditions case reaches 34 km s$^{-1}$ at $T=$ 5.37 MK and 60 km s$^{-1}$ at $T=$ 7.76 MK, while for the high initial conditions case $\xi=0$ at both temperatures. This example shows dramatically the effect that initial conditions have on the line width.

The previous analysis shows how a diversity of nanoflare energies, frequencies and initial conditions (as shown in Figures~\ref{fig:nontherm_nano} and~\ref{fig:nontherm_init_cond}) lead to the variety of non-thermal velocities obtained from the 2DCAM-EBTEL samplings of Figure~\ref{fig:nontherm_loop}.


\section{Conclusions}
\label{sect:conclusions}

We studied the effect that the plasma evolution produced by a nanoflare model of coronal heating has on the characteristics of EUV spectral lines. We used the 2DCAM-EBTEL model to produce a series of evolving strands whose emission was integrated to create synthetic lines. We analyzed these lines by applying common plasma diagnostic techniques. From the modeled lines we computed Doppler-shift velocities, line asymmetries and non-thermal widths as it is usually done with observed spectral lines. We found a series of characteristics of the modeled lines that can be compared with actual spectral observations and could be used to guide future investigations. 

The diverse properties of the analyzed lines are due to a complex combination of different plasma temperatures, densities and velocities in simultaneously evolving strands. As our model considers the transition region and coronal emissions separately, we could also compare the relative weight of each contribution at different temperatures and the role they play in defining some of the line characteristics. 

Regarding the dependence of Doppler-shift velocities on temperature, we found downflows of a few km s$^{-1}$ for temperatures below 4 MK and upflows up to 20 km s$^{-1}$, with a mean of 14 km s$^{-1}$, for the hotter analyzed line corresponding to the FeXIX ion, at a formation temperature of 7.76 MK (see Table~\ref{table1} and Figure~\ref{fig:dop_vel_loop}a). Some of these velocities are consistent with observed values obtained by other authors \citep{warren2011b,winebarger2013,tripathi2012a}, but only up to a certain point, since observational works found a shift from downflows to upflows at temperatures that are lower ($\sim$ 1 MK) than those predicted by our model ($\sim$ 3 MK) \citep[see also,][]{tripathi2012b,peter1999}. This apparently high temperature for the switch from downflows to upflows relates both, to the relative weight of the TR and coronal emission combined with the width of the line contribution function, but also to the way in which the TR and coronal temperatures and velocities are linked. As described in Section~\ref{sect:lines} the TR temperature can be as high as 60\% of the coronal temperature. At usual peak nanoflare temperatures above 5 MK, TR temperatures in the model can be as high as 3 MK. As we have seen (see CaXIV line in Table~\ref{table1}), at those temperatures the coronal and TR contributions are comparable. A similar link between the TR and coronal temperatures and the presence of high temperature downflows has been identified by \citet{testa2016} in their study using a Bifrost 3D MHD simulation to guide the interpretation of their observations.

All models tend to overestimate the brightness of the transition region relative to the corona. \citet{warren2010} suggested that this may be because the models do not include the rapid expansion of the field as it emerges from intense magnetic elements in the photosphere and rapidly flares out. If the transition region is in the throat of these flux tubes, its area will be less than that in the corona a short distance above \citep[though see][]{guarrasi2014}. Whatever the cause, it is likely that the transition region in our model is too bright \citep{schonfeld2020}. This would push the crossover between red and blue shift to a higher temperature than observed, consistent with what we find.

The coronal velocity we report is the velocity at the base of the corona (top of the transition region). The flow must decelerate to zero at the strand apex under our assumed symmetric conditions. In addition, we assume that the line of sight is always along the strand axis, which is not the case for an arching structure. Both effects decrease the average coronal Doppler shift relative to the values we use.

It is also worth to note that our results assume that the transition region and corona are observed together, which may not be true for observations. Many lines-of-sight in an active region pass only through the corona, missing the transition region. The transition region appears as moss, and observations of inter-moss regions do not include transition region emission. On the other hand, lines-of-sight that do include moss likely give excess weighting to the transition region. Our results are best compared against observations that average over an active region.

Clearly, further modeling and a more thorough comparison with observations are needed to help steer future investigations. The analysis of lines integrated from the evolution of individual nanoflares with different characteristics helped us understand what features of the plasma evolution are behind the collective effect of several evolving strands on the modeled spectral lines (see the discussions of Section~\ref{sect:doppler}).

Another feature of spectral lines usually analyzed in plasma diagnostics is the non-thermal broadening. Our results show that non-thermal velocities between approximately 9 and 16 km s$^{-1}$, with little dispersion, are expected up to 3 MK (see Table~\ref{table1} and Figure~\ref{fig:nontherm_loop}a). Similar non-thermal velocity values in the temperature range from 1 to 4 MK have been previously reported in active region observations by other authors \citep[see e.g., ][]{hara1999,brooks2016,testa2016}, although perhaps with somehow larger non-thermal velocities and standard deviations. For the hotter lines, we obtain larger dispersions in the non-thermal velocities, with a variation from 0 to 20 km s$^{-1}$ and a mean of 12 km s$^{-1}$ for the FeXVII line at 5.37 MK, and a variation from 0 to 50 km s$^{-1}$ with a mean of 30 km s$^{-1}$ for FeXIX line at 7.76 MK (see Figure~\ref{fig:nontherm_loop}a). 

There are sources of flow other than the evaporation and draining of a nanoflare cycle that we consider here. Both magnetic reconnection and waves – the sources of the nanoflare energy – produce high velocity motions perpendicular to the magnetic field, especially at high temperature. These would be manifested primarily as nonthermal line broadening \citep{cargill1996}. 

Because of all the above factors, our results only provide general guidance on the ability of nanoflares to reproduce observed Doppler shifts and line broadening. The overall agreement is satisfactory. We have shown that Doppler shift and broadening depend on nanoflare parameters to an extent that they may provide meaningful diagnostics. Rigorous tests must await more detailed modeling of the type we are planning for the future and a more targeted comparison with observations. New observations expected from upcoming missions will indeed be very helpful.

Although recent observations provided evidence of the presence of a hot plasma component that supports the idea of nanoflare heating \citep[see e.g.,][and references therein]{brosius2014, ishikawa2017,hinode-team2019}, spectroscopic observations at temperatures above 5 MK have been sparse. One of the reasons is that the very weak emission at those temperatures is not easily picked up by available instruments. Presently, there is in development a new generation of high spectral resolution instruments sensitive to very hot plasmas, such as the multi-slit EUV spectrometer on board the Multi-slit Solar Explorer \citep[MUSE, ][]{depontieu2020,depontieu2022,cheung2022} and the EUV High-Throughput Spectroscopic Telescope \citep[EUVST, ][]{shimizu2019} on board Solar-C. Also, a soft X-ray spectrometer in the 90-150 \AA~range has been recently proposed \citep[see ][]{delzanna2021}. These instruments would provide more definitive answers regarding the presence of mass motions at very high temperatures which is a typical signature of nanoflare heating.

The effort to study the effect that different heating mechanisms and regimes have on coronal spectral lines is not new. In recent years different groups have investigated the problem by means of modeling and observations, among these: Joule dissipation in 3D magnetohydrodynamic models of the corona \citep{peter2006,hansteen2010}; nanoflare hydrodynamic modeling \citep{patsourakos2006}; nanoflare heating by electron beam injection produced by reconnection \citep{testa2014,testa2016,polito2018}; braiding turbulence \citep{pontin2020} and Alfv\'en wave turbulence models \citep{asgari-targhi2014}. A thorough review can be found in \citet{depontieu2022}. The novelty in the present work is the use of a simple model (see Section~\ref{sect:model}) based on the simultaneous evolution of several strands being heated by nanoflares. Perhaps one of the most important results obtained here is the prediction of high Doppler (upflow) and non-thermal velocities in the higher temperature range ($T>$ 5 MK). Although the ultimate goal of the kind of analysis presented here is to contribute to understand the causes of coronal heating, we do not pretend that the obtained results would provide definitive answers. On the contrary, we hope that our findings and their comparison with actual observations would help to guide future research.
 
Our choice of 2DCAM-EBTEL to explore the effect of nanoflare heating on spectral lines was based on the success of the model to reproduce other observed features of active region plasmas, such as the statistical properties of lightcurves in different wavelengths and emission measure distributions (see LFK15 and LFK16). One advantage of the model is EBTEL's low demand of computational power, which allows one to run the evolution of many strands in reasonable times with a modest computer. The density of strands per cross magnetic-field surface unit, is still one of the basic unknowns of the nanoflare model \citep{pontin2020,williams2021}. The number considered here (49 strands) proved to be a proper value according to our previous works and it is also convenient given the statistical nature of the present study.

It is worth noting that despite its advantages in usability over more numerically demanding codes, EBTEL is based in a number of approximations and assumptions. Therefore, it will be important to validate some of the results found here by comparing EBTEL's output with more sophisticated models, particularly in extreme conditions during the plasma evolution in which its approximations may be prone to fail \citep[see e.g., ][]{rajhans2021}.

One of the approximations usually made both in models and plasma diagnostics is that of equilibrium ionization. However, it has been shown that non-equilibrium ionization during the evolution of rapidly heating plasmas and related flows, and specially for very hot lines in low-frequency heating situations, can have a considerable effect in the ion populations and line intensities \citep[see e.g., ][]{imada2011,dudik2017,bradshaw2011}. This will be analyzed in future work by comparing the results obtained here with a model that addresses these problems more properly like the 1D Hydrodynamic and Radiation code \citep[HYDRAD, ][]{bradshaw2013}.

\begin{acknowledgements}

The authors thank the anonymous referee for fruitful comments and suggestions. M.L.F. is a member of the Carrera del Investigador Cient\'{\i}fico of the Consejo Nacional de Investigaciones Cient\'{\i}ficas y T\'ecnicas (CONICET). M.L.F. acknowledges financial support from the Argentinean grants PICT 2020-03214 (ANPCyT) and PIP 11220200100985CO (CONICET). J.A.K.’s effort was supported by the GSFC Internal Scientist Funding Model (competitive work package program).

\end{acknowledgements}



\bibliography{flows_paper}{}
\bibliographystyle{aasjournal}

\end{document}